\documentclass[fleqn,usenatbib]{mnras}

\usepackage{newtxtext,newtxmath}

\usepackage[T1]{fontenc}
\usepackage{float}

\usepackage{comment}
\usepackage{ulem}
\DeclareRobustCommand{\VAN}[3]{#2}
\let\VANthebibliography\thebibliography
\def\thebibliography{\DeclareRobustCommand{\VAN}[3]{##3}\VANthebibliography}

\usepackage{graphicx}	
\usepackage{amsmath}
\hypersetup{
     colorlinks = true,
     linkcolor = blue,
     anchorcolor = blue,
     citecolor = blue,
     filecolor = blue,
     urlcolor = blue
}
\usepackage{soul}

\title[Pulsar's time delay at the GC]{Testing space-time geometries and theories of gravity at the Galactic Center with pulsar's time delay}

\author[R. Della Monica et al.]{
Riccardo Della Monica,$^{1}$\thanks{E-mail: rdellamonica@usal.es}
Ivan De Martino,$^{1,2}$\thanks{E-mail: ivan.demartino@usal.es}
Mariafelicia De Laurentis,$^{3,4}$\thanks{E-mail: mariafelicia.delaurentis@unina.it}
\\
$^{1}$Universidad de Salamanca, Departamento de Fisica Fundamental, P. de la Merced, E-37008 Salamanca, Spain\\
$^{2}$  Instituto Universitario de Física Fundamental y Matemáticas (IUFFyM), P. de la Merced, E-37008 Salamanca, Spain\\
$^{3}$Dipartimento di Fisica, Universit\'a
di Napoli {}``Federico II'', Compl. Univ. di
Monte S. Angelo, Edificio G, Via Cinthia, I-80126, Napoli, Italy\\
$^{4}$ INFN Sezione  di Napoli, Compl. Univ. di
Monte S. Angelo, Edificio G, Via Cinthia, I-80126, Napoli, Italy
}

\date{Accepted 2023 Jul 12. Received 2023 Jul 12; in original form 2023 May 31}

\pubyear{2023}

\begin{document}
\label{firstpage}
\pagerange{\pageref{firstpage}--\pageref{lastpage}}
\maketitle

\begin{abstract}
    We developed a numerical methodology to compute the fully-relativistic propagation time of photons emitted by a pulsar in orbit around a massive compact object, like the supermassive black hole Sagittarius A* in the Galactic Center, whose gravitational field is described by a generic spherically symmetric space-time. Pulsars at the Galactic Center are usually regarded as the next major precision probe for theories of gravity, filling the current experimental gap between horizon-scale gravity tests and those at larger scales. We retain a completely general approach, which allows us to apply our code to the Schwarzschild space-time (by which we successfully validate our methodology) and to three different well-motivated alternatives to the standard black hole 
    paradigm. The results of our calculations highlight departures spanning several orders of magnitudes in timing residuals, that are supposed to be detectable with future observing facilities like the Square Kilometer Array.
\end{abstract}




\section{Introduction}

After more than a century since its original formulation by A. Einstein, the theory of General Relativity (GR) has survived all the experimental tests with flying colors \citep{Will2018}. While the so-called ``classical tests of GR'' provided early confirmations of the theory in our Solar System, observational pieces of evidence have grown in number and complexity over the years, providing experimental validations of the predictions from GR in plenty of different astrophysical scenarios \citep{Will1993,Will2014,Turyshev2009,Tasson2016,Hees2016}. Recently, thanks to the technical and theoretical advancement that led to the first direct detection of gravitational waves from mergers of compact objects \citep{Abbott2016a,Abbott2016b}, to the imaging of the shadow of supermassive black holes (SMBHs, \citet{EventHorizonTelescopeCollaboration2019a,Akiyama2022a}), and to the detection of general relativistic effects on the orbits of stars in the Galactic Center (GC, \citet{GravityCollaboration2018a,Do2019a,GravityCollaboration2020c}) of the Milky Way (MW), new avenues to test GR in the strong-field regime have been opened. All such tests have increasingly narrowed down the margin for possible deviations from GR (we refer  to \citet{deLaurentis2023} for a comprehensive review). Nonetheless, the pursuit for observational signatures that could provide a smoking gun for alternative theories of gravity has not been diminished. On the other hand, such new opportunities to develop and carry out tests for alternatives to GR with unprecedented precision \citep{Berti2015} reignited the field.

\begin{figure}
    \centering
\includegraphics[width = 1.0\columnwidth]{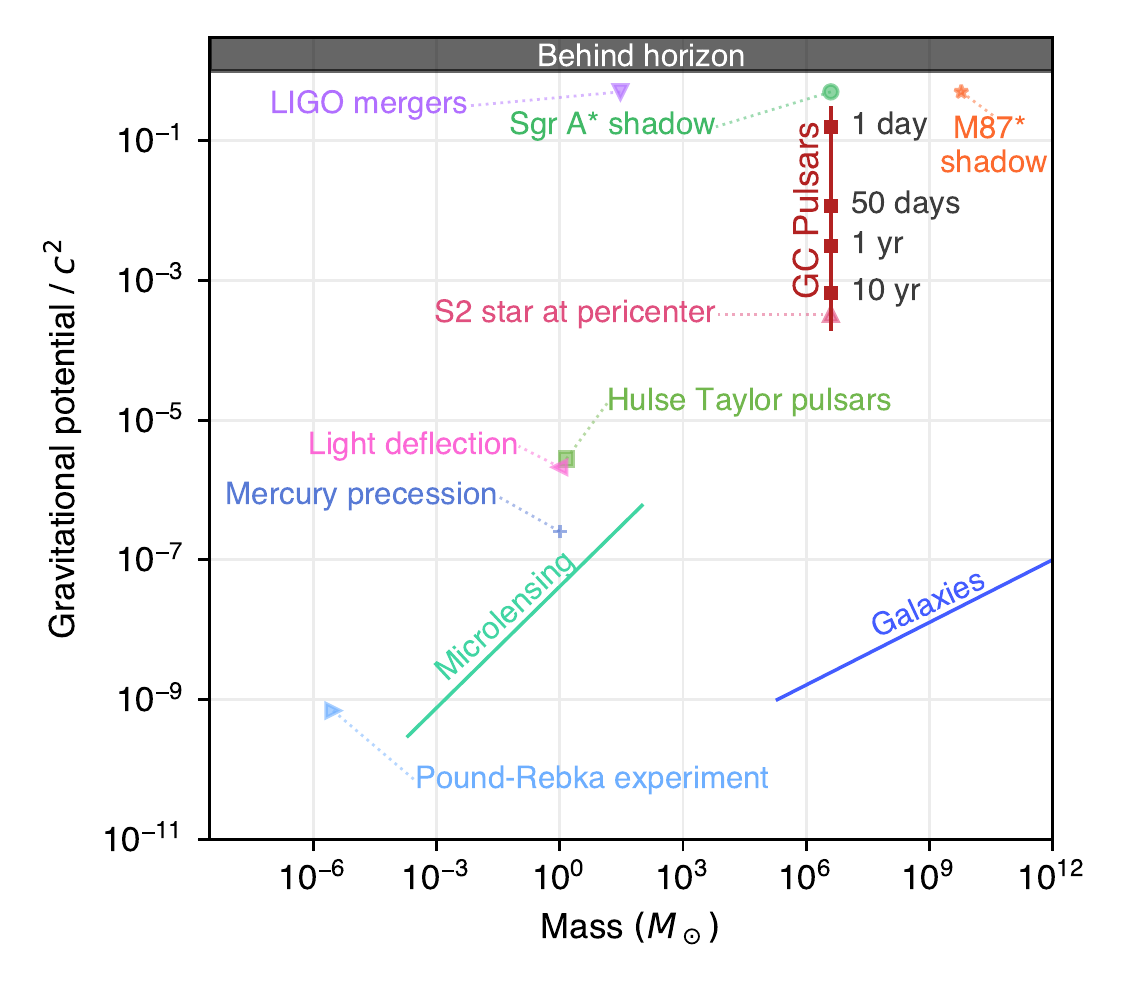}
    \caption{The gravitational potential for different astrophysical probes of gravity, as a function of the mass of the gravitational source. Pulsars at the GC would allow to fill the experimental gap between the S-stars analyses and horizon scales tests of gravity.}
    \label{fig:observational_regimes}
\end{figure}

The next major breakthrough in experimental gravitation is expected to come from the discovery of pulsars orbiting a SMBH. Thanks to their astounding intrinsic rotation stability (typically, variations are on the order of one part in $10^{15}$ per pulse period \citep{Becker2018}), pulsars are considered among the best tools to probe gravitational fields \citep{Stairs2003,Lorimer2008,Will2014}. Starting from the 1970s, the discovery and study of double-pulsars systems have allowed several general relativistic effects to be detected with increasingly high sensitivity, such as the orbital period decay due to dipolar gravitational waves emission \citep{Hulse1975,Taylor1994,Kramer2006a}. These systems, however, probe a comparatively weak gravitational field, with masses of the components of the system ($M$) on the order of units of solar masses and compactness $GM/Rc^2\sim10^{-5}\div {10}^{-7}$ \citep{Zhang2017a}, being $R$ the binary separation. However, a pulsar on a tight orbit (with orbital periods below 1-100 yrs) around a supermassive compact object like Sagittarius A* (Sgr A*, $M\sim 4\times {10}^6 M_\odot$) in our GC, would probe a totally different regime of gravity (see Fig. \ref{fig:observational_regimes}), and analysis of the times-of-arrival (TOA) of the pulses emitted by such objects (the so-called \textit{pulsar timing analysis}) would supersede all previous tests of GR in the strong field regime \citep{Wex1999,Liu2012}. Besides an unparalleled improvement in the mass determination of Sgr A*, pulsar timing analyses in the GC would allow the measurement of the spin magnitude and orientation of the SMBH \citep{Liu2012,Zhang2017a} with a precision of order ${10}^{-4}\div {10}^{-3}$, the measurement of the SMBH quadrupole moment \citep{Wex1999,Liu2012,Psaltis2016} 
and thus a direct test of the no-hair theorem at the GC \citep{Christian2015, Izmailov2019}. Furthermore, relativistic effects on both the pulsar's trajectory and the pulses photon paths would set the stage for unprecedented metric tests of the space-time geometry around a SMBH, allowing to extend the already existing tests of alternatives to the standard BH paradigm 
\citep{DeLaurentisY2018,DeMartino2021,DellaMonica2022a,DellaMonica2022b,DellaMonica2023c,Cadoni2023} to a much higher expected level of precision.

The great scientific potential of pulsars at the GC has motivated a growing number of radio pulsar searches within the central few parsecs of the MW \citep{Johnston1995,Johnston2006,Deneva2009,Deneva2010, Bates2011}. Despite the efforts, all such searches have basically failed in finding the pulsar population they sought after, with only six pulsars  discovered within 15 arcmins of Sgr A* \citep{Deneva2009} and only one radio magnetar 2.4 arcsec (0.1 parsecs in projection) away from Sgr A* \citep{Kennea2013,Mori2013,Rea2013}.
This remarkable scarcity of observed GC pulsars is believed to be resulting from interstellar scattering processes whose effect is a temporal broadening of the pulses as they pass through the heavily turbulent and ionized interstellar medium within the GC \citep{Cordes2002}. This process has a strong dependence ($\propto\nu^{-4}$) on the observing frequency $\nu$, making the usual periodicity search techniques at frequencies $\nu \sim 10^9$ Hz basically useless, even for long-period pulsars. Pulse temporal broadening cannot be compensated or corrected by instrumental means \citep{Eatough2013a}, so the only potential way to alleviate the problem is to move pulsar searches to higher observing frequencies. Unfortunately, due to the characteristic power-law spectra of pulsars $\propto \nu^{\alpha}$ (with $\alpha < 0$, \citet{Wharton2012}), a higher observing frequency corresponds to a dimmer intrinsic source flux. For this reason, all the high-frequency pulsar searches that have been performed so far in the GC (even the most recent search at 2 and 3 mm, \citet{Torne2021}) failed to report the detection of new pulsars therein. However, the presence of a numerous population of young and massive stars orbiting Sgr A* \citep{Paumard2006, Lu2013} suggests that pulsars originating from supernovae explosions of the massive end ($> 9 M_\odot$) of such a population should indeed be there. Moreover, depending on the specific population model, it is estimated that between 100 and 1000 pulsars should reside within the central parsec of the GC with orbital period $<100$ yr (with around 100 pulsars with orbital periods $<10$ yr) and more than 10000 millisecond pulsars \citep{Pfahl2004,Zhang2014,Rajwade2017,Chennamangalam2014}.

For this reason, the detection of at least one pulsar on a tight orbit around Sgr A* is a major scientific goal of future observational facilities, like the Square Kilometre Array (SKA, \citet{Keane2015}), the Five-hundred-meter Aperture Spherical Telescope (FAST, \citet{Nan2011}), the next generation Very Large Array (ngVLA, \citet{Bower2018}) or the Event Horizon Telescope \citep{EventHorizonTelescopeCollaboration2022a}. Due to their large collection areas, such facilities expect not only to detect pulsars around Sgr A* but also to be able to perform timing analysis \citep{Eatough2015}.

In this work, we envision, develop, and test a numerical methodology to compute the photon propagation time of the pulses emitted by a pulsar in a generic orbit around a massive compact object, whose gravitational field is described by a spherically symmetric space-time. We built a code to integrate geodesic equations for the motion of the pulsars (treated as a test particle due to the extreme mass ratio with the SMBH), solve the emitter-observer problem for the null geodesic connecting the emitting pulsar and a distant observer and integrate the propagation time. This approach takes into consideration all the GR effects on the orbit and on the photon, without resorting to post-Newtonian or post-Keplerian approximations \citep{Damour1986}. 
The emitter-observer problem in the case of a pulsar orbiting a Schwarzschild black hole was previously investigated by \citet{Hackmann2019} limiting the study to the case of a pulsar on a circular orbit. Moreover, a fully-relativistic numerical treatment of pulsar timing around a spinning SMBH in the GC, described by the Kerr metric \citep{Kerr1963}, has been first approached in \citet{Zhang2017a} and in more recent works \citep{Kimpson2019} with a focus on the radio timing of potential SMBH-orbiting millisecond pulsars \citep{Kimpson2020a, Kimpson2020b} . Here, we extend the fully-relativistic approach to a generic spherically symmetric space-time. In such a way, we construct a model-independent approach that is able to test wide classes of stationary black hole solutions coming from different theories of gravity, and different metrics describing black hole mimickers in GR.  
The outline of the paper is as follows: Section \ref{sec:photon_propagation_time} is devoted to an overview of the classical treatment for computing the photon propagation time in a Schwarzschild space-time, either by post-Newtonian formulae or with an exact fully-relativistic approach; in Section \ref{sec:numerical} we describe the technical details of our code for a generic spherically symmetric space-time; in Section \ref{sec:alternatives} we test our methodology for a Schwarzschild space-time and compare the numerical results with the exact formula provided by \citet{Hackmann2019}, and we apply it to three alternatives to the classical BH paradigm; Section \ref{sec:results} reports our results, while we report our conclusions in Section \ref{sec:conclusions}.

\section{An overview on the photon propagation time}
\label{sec:photon_propagation_time}

The problem of pulsar timing has been historically formulated for binary pulsar systems, in order to develop techniques able to fit the increasing amount of TOA data for such systems found in the local Universe \citep{Hulse1975, Demorest2010, Antoniadis2013, Cromartie2020, Fonseca2021} and due to the importance that they assumed in the field of experimental gravitation \citep{Will1977, Verbiest2008, Kramer2021a}. Our understanding and our ability to model the time delay of double pulsar systems heavily rely on the post-Newtonian approximation \citep{Will1977, Damour1983, Damour1986, Damour1988a, Damour1988b, Damour1988c} by which the general relativistic motion of such a system is described with remarkable accuracy. At the post-Newtonian level, one can consider, 
perturbatively, all sorts of relativistic effects, both on the pulsars motion and on the propagation of light rays, up to any order of expansion, without having to solve the fully relativistic two body problem by which the dynamics of the system is regulated and for which no closed form solution can be formulated \citep{Damour2013}. A comprehensive treatment of all possible sources of delay and their expression in the post-Newtonian approximation can be found in the pioneering work by \citet{Damour1986}, on whose basis all modern timing codes are formulated \citep{Edwards2006}.

For our purpose, we will focus on the first-order post-Newtonian effects that, due to the perturbative nature of the post-Newtonian approximation, represent the dominant contribution to the timing delay and we will particularise it for pulsars orbiting SMBHs. First of all, we need to distinguish the effects that change the position and the time at which photons are emitted by the pulsar with respect to the Newtonian case (\textit{e.g.} the orbital precession and the Einstein delay), from those that directly alter the photon travel time (the Rømer delay and the Shapiro delay). 
More specifically, differently from Keplerian orbits, trajectories in GR (and eventually in modified theories of gravity) do not coincide with closed ellipses but suffer from pericenter advance that, on each orbital period, shifts the angular position of the pericenter by an angle
\begin{equation}
    \Delta\omega_{\rm GR} = \frac{6\pi GM}{c^2a(1-e^2)},
    \label{eq:gr_precession}
\end{equation}
being $M$ the mass of the central SMBH, $a$ the semi-major axis of the pulsar's orbit and $e$ its eccentricity. Clearly, this post-Keplerian effect changes the position in space, $\vec{r}_{\rm e}$, from where the photon is emitted by the pulsar. Additionally, relativistic effects at first post-Newtonian order can also alter the time of the emission $t_{\rm e}$ as perceived by a distant observer. This shift is related to a slow-down of the pulsar's proper time with respect to the coordinate time measured by such an observer, due to a combination of special relativistic and gravitational time dilation \citep{Damour1986,  Blanchet2001, Poisson2014}. These contributions sum up to the so-called Einstein delay whose amplitude, for a pulsar-SMBH system, is given by \citep{Blandford1976, Liu2014}:
\begin{equation}
    \gamma_{\rm E} = \frac{2e}{c^2}\left(\frac{(GM)^2T}{2\pi}\right)^{1/3},
\end{equation}
where $T$ is the pulsar's orbital period around the SMBH. This amplitude is modulated along the orbit according to the law
\begin{equation}
    \Delta t_{\rm E} = \gamma_{\rm E}\sin u,
    \label{eq:einstein_delay}
\end{equation}
where $u$ is a parameter corresponding to the orbital eccentric-anomaly (we refer to \citet{Damour1986} for more details).

On the other hand, at first order, one can consider the propagation of photons on a straight line \citep{Will2014} and compute their travel time from the emitter's position $\vec{r}_{\rm e}$ to the observer's position $\vec{r}_{\rm o}$ as a linear sum of different effect:
\begin{equation}
    \Delta t_{\rm PN} = \Delta t_{\rm R}+\Delta t_{\rm  Sh}.
    \label{eq:post_newtonian}
\end{equation}
Here, $\Delta t_{\rm R}$ represents the classical Rømer delay related to the photon propagation time across the pulsar's orbit
\begin{equation}
    \Delta t_{\rm R}=\frac{|\vec{r}_{\rm o}-\vec{r}_{\rm e}|}{c}\,,
\end{equation}
and $\Delta t_{\rm Sh}$ is the Shapiro time delay related to the time dilation experienced by light rays when grazing the region where the central object curves space-time substantially, and is given by:
\begin{equation}
    \Delta t_{\rm Sh}=-\frac{2GM}{c^2}\ln\left(\frac{2|\vec{r}_{\rm o}|}{|\vec{r}_{\rm e}|+\vec{r}_{\rm e}\cdot\vec{n}}\right)\,,
    \label{eq:shapiro}
\end{equation}
where 
$\hat{n} \equiv \vec{r}_{\rm o}/|\vec{r}_{\rm o}|$ is the unit vector pointing from the emitter to the observer. It is worth noticing that both formulas for the Rømer and the Shapiro delay are computed by assuming that the photons propagate on a straight line. As mentioned, the advantage of a post-Newtonian approach is that one doesn't have to tackle the problem of solving the fully relativistic equations of motion for neither the pulsar nor the photons.
However, differently from the double-pulsar settings, where the mass of the two companions are generally comparable with each other (thus requiring solving a full two-body problem, which doesn't generally have a closed-form solution), in the case of a pulsar orbiting a SMBH the extreme mass ratio allows to effectively treat the pulsar as a test particle in the gravitational field of the massive object. This consideration opens to the possibility of approaching the delay problem with a completely analytical treatment, without resorting to approximations of any sort. 

\subsection{The photon propagation time around a Schwarzschild black hole}

The advantage of retaining a fully relativistic approach lies in the ability to describe strong relativistic effects (both on the pulsar orbit and on the photon propagation) in a self-consistent way without the need to introduce the post-Keplerian and post-Newtonian approximations.
This is the approach adopted by \citet{Hackmann2019} for the case of a Schwarzschild space-time describing the gravitational field of a point mass $M$ in GR,
\begin{align}
    ds^2 &= g_{\mu\nu}dx^\mu dx^\nu=\nonumber \\&=-\left(1-\frac{2M}{r}\right)dt^2+\left(1-\frac{2M}{r}\right)^{-1}dr^2+r^2d\Omega^2,
    \label{eq:schwarzschild_spacetime}
\end{align}
where $(t,\,r,\,\theta,\,\phi)$ are the usual Schwarzschild coordinates \citep{Wald1984} and $d\Omega^2 = d\theta^2+\sin^2\theta d\phi^2$ is the angle element on a sphere and the signature $(-,+,+,+)$ is adopted. 
The relativistic equations of motion for a test particle in this gravitational field are defined as the geodesic equation related to the metric in Eq. \eqref{eq:schwarzschild_spacetime},
\begin{equation}
    \frac{d^2x^\mu}{d\lambda}+\Gamma^\mu_{\nu\rho}\frac{dx^\nu}{d\lambda}\frac{dx^\rho}{d\lambda} = 0,
    \label{eq:geodesic_eq}
\end{equation}
where $\lambda$ is an affine parameter on the geodesic (\textit{i.e.} the proper time in the case of a massive particle) and $\Gamma^\mu_{\nu\rho}$ are the Christoffel symbols built from the metric coefficients. Solving Eq. \eqref{eq:geodesic_eq} for a time-like geodesic ($g_{\mu\nu}\dot{x}^\mu\dot{x}^\nu = -1$, where a dot represents a derivative with respect to the affine parameter $\lambda$) allows to describe the dynamics of massive test particles in the gravitational field in Eq. \eqref{eq:schwarzschild_spacetime},
taking into account, in a self-consistent way, all the orbital relativistic effects (including the orbital precession in Eq. \eqref{eq:gr_precession}, both special and general relativistic time dilation in Eq. \eqref{eq:einstein_delay} and higher order orbital perturbations) without having to resort to the approximated post-Newtonian expressions.
On the other hand, particularizing the geodesic equations for a null geodesic allows describing the motion of massless particles (\textit{i.e.} photons) in the space-time described by Eq. \eqref{eq:schwarzschild_spacetime}. Starting from there, the photon travel the time can be written as the integral
\begin{equation}
    c\Delta t = \int_\gamma \frac{r^2 dr}{b\left(1-\frac{2M}{r}\right)\sqrt{R(r)}}\,,
    \label{eq:travel_time_integral}
\end{equation}
depending on the photon path $\gamma$ connecting emitter and observer, on the corresponding impact parameter $b$ and on the fourth-degree polynomial function $R(r) = r^4/b^2-r^2+2Mr$. 
In Schwarzschild space-time, the integral in Eq. \eqref{eq:travel_time_integral} has an exact solution given by
\begin{equation}
    \Delta t_{\rm ex} = \frac{GM}{c^3}\left(T(r_{\rm{o}}, b)\pm T(r_{\rm e},b)\right).
    \label{eq:exact_formula}
\end{equation}
The sign $\pm$ in Eq. \eqref{eq:exact_formula} depends on whether the integration path is direct or indirect (see Sec. \ref{sec:numerical} for a more detailed definition) and the functions $T(r,b)$ are, for a specific impact parameter, the results of the integral in Eq. \eqref{eq:travel_time_integral}, given analytically by \citet{Hackmann2019}:
\begin{align}
    T(r,b) =& \frac{2}{\sqrt{r_4(r_3-r_1)}}(T_1+T_2+T_3+T_4) + T_\infty\,,
\end{align}
where we have defined the terms
\begin{align}
    T_1 &= \left(\frac{r_3^3}{r-2}+\frac{1}{2}(r_4-r_3)(r_3-r_1+4)\right)F(x,k),\\
    T_2 &= -\frac{1}{2}r_4(r_3-r_1)E(x,k),\\
    T_3 &= -2(r_4-r_3)\Pi\left(x,\frac{k^2}{c_1},k\right),\\
    T_4 &= -\frac{8(r_4-r_3)}{(r_4-2)(r_3-2)}\Pi(x,c_2,k)\,,
\end{align}
and $T_\infty$ encodes all the diverging (for $r\to\infty$) terms as follows
\begin{equation}
    T_\infty = \frac{b\sqrt{R(r)}}{r-r_3}+2\ln\left(\frac{\sqrt{r(r-r_1)}+\sqrt{(r-r_4)(r-r_3)}}{\sqrt{r(r-r_1)}-\sqrt{(r-r_4)(r-r_3)}}\right).
\end{equation}
In the previous relations, all quantities that appear are dimensionless (\textit{e.g.} $r$ and $b$ are expressed in units of gravitational radii $GM/c^2$), the functions $F$, $E$ and $\Pi$ are Jacobian elliptic integrals \citep{NISTDLMF}, the quantities $r_1$, $r_2$, $r_3$, $r_4$ are the roots of the polynomial $R(r)$ in Eq. \ref{eq:travel_time_integral}, $x$ is an auxiliary variable and $k$, $c_1$ and $c_2$ are all constant terms built from the roots $r_1$, $r_2$, $r_3$, $r_4$ as shown in Appendix A of \citet{Hackmann2019}. This timing formula has been shown to provide a better description of the photon propagation time with respect to the usual post-Newtonian approximations, being able to capture strong field features that the first order post-Newtonian in Eq. \eqref{eq:post_newtonian} effect fail to capture. These can amount to deviations on the order of seconds when considering a circular orbit with a radius of about a hundred gravitational radii.

\section{Numerically solving the emitter-observer problem in a spherically symmetric space-time}
\label{sec:numerical}

\begin{figure*}
    \centering
    \includegraphics[width=\textwidth]{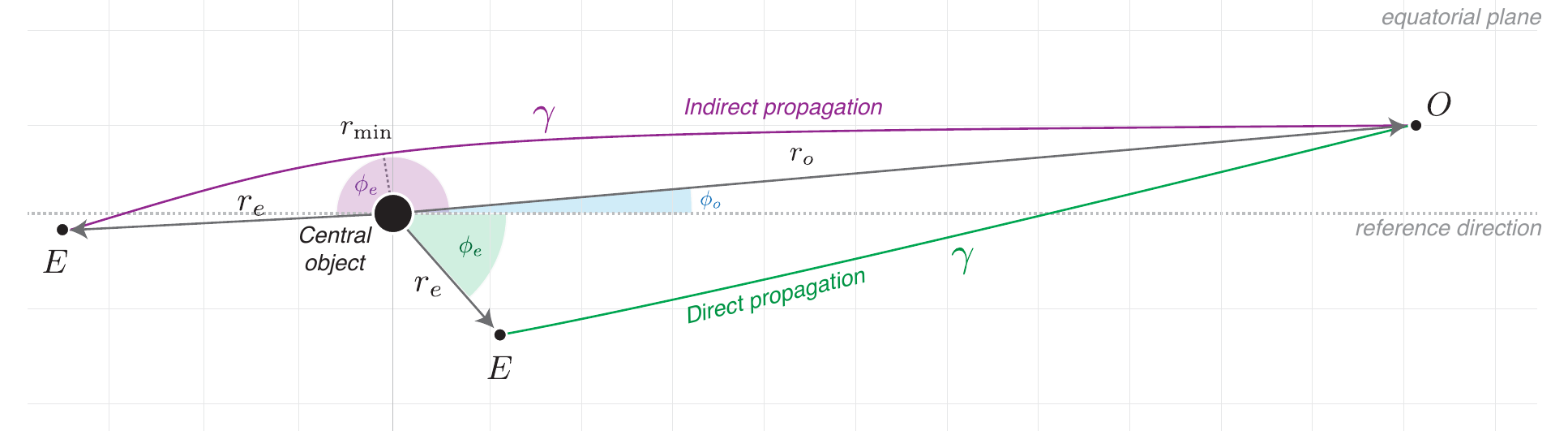}
    \caption{Illustration of the configuration for the emitter-observer problem. The emitter is located at a point $E$ that is identified by polar coordinates $(r_{\rm e},\phi_{\rm e})$, while the observer receiving the photon is located at point $O$ with coordinates $(r_{\rm o} \phi_{\rm o})$. Considering only primary photons received by the observer (\textit{i.e.} we do not consider photons that graze so close to the unstable photon orbit of the central object that their paths bend so  strongly, $\Delta \phi > 2\pi$, that they reach the observer after one or more complete turns around the central object) only two possible scenarios are possible: \textit{(green path)} the radial coordinate increases monotonically going from $r_{\rm e}$ to $r_{\rm o}$ \textit{propagating directly} ($r_{\rm e}\to r_{\rm o}$) from the emitter to the observer; \textit{(purple path)} the photon leaves the observer with a decreasing radial coordinate (\textit{i.e.} a negative radial velocity) then reaches a minimum distance $r_{\rm min}$ from the central object after which it starts increasing again up to the observer position. We call the latter configuration \textit{indirect propagation} ($r_{\rm e}\to r_{\rm min} \to r_{\rm o}$).}
    \label{fig:emitter_observer}
\end{figure*}

In Eq. \eqref{eq:exact_formula}, the photon propagation time is reported for the light rays emitted by a pulsar in orbit around a Schwarzschild BH that travel up to a distant observer, as derived in \citet{Hackmann2019}. Despite the possibility of solving this problem analytically, it is clear from Eq. \eqref{eq:exact_formula} that the specific expression derived in  \citet{Hackmann2019} is only valid under the assumption of a Schwarzschild space-time geometry given in Eq. \eqref{eq:schwarzschild_spacetime}. It is reasonable to assume that once one takes into consideration models whose geometry differs from that of a Schwarzschild BH (see \textit{e.g.} \citeauthor{EventHorizonTelescopeCollaboration2022f} \citeyear{EventHorizonTelescopeCollaboration2022f}, \citet{Vagnozzi2022} and \citet{deLaurentis2023} for an overview of the plethora of such possible alternatives), such an expression should be modified accordingly for each model and might even result in the impossibility of solving the corresponding integrals analytically.  
For this reason, below we will approach the problem by developing a numerical methodology for the computation of the photon propagation time regardless of the specific model considered. For this purpose, let's consider a generic asymptotically-flat spherically-symmetric space-time described, in the usual Schwarzschild coordinates $(t,\,r,\,\theta,\,\phi)$, by the line element
\begin{equation}
    ds^2 = -A(r)dt^2+B(r)dr^2+r^2d\Omega^2,
    \label{eq:generic_metric}
\end{equation}
where $d\Omega^2 = d\theta^2+\sin^2\theta d\phi^2$. We are interested in the computation of the propagation time delay for photons following null geodesic-paths in such space-time. Retracing the same line of reasoning presented in \citet{Hackmann2019}, one can restrict (without loss of generality, due to the spherical symmetry of the problem) to the equatorial plane $\theta = \pi/2$ (which may not coincide with the plane on which the orbit of the emitting object lies), and consider the propagation of light rays on this plane. One can than define the conserved specific (\textit{i.e.} per unit mass of the test particle) 
energy, $\mathcal{E} = -{\partial\mathcal{L}}/{\partial t}$, and angular momentum, $L ={\partial\mathcal{L}}/{\partial \phi}$, for a null geodesic, where $\mathcal{L} = g_{\mu\nu}\dot{x}^\mu\dot{x}^\nu$ is the Lagrangian of a test particle.
This quantity is itself a constant of motion as its value, for a null geodesic, is identically zero due to the normalization of the 4-velocity ($g_{\mu\nu}\dot{x}^\mu\dot{x}^\nu = 0$). One can then define from the conserved quantities the impact parameter $b\equiv {L}/\mathcal{E}$ 
which allows rewriting the equations of motion for a null geodesic as
\begin{align}
    &\frac{dt}{d\tau} = \frac{\mathcal{E}}{A(r)},\label{eq:dtdtau}\\
    &\frac{d\phi}{d\tau} = \frac{L}{r^2},\label{eq:dphidtau}\\
    &\frac{dr}{d\tau} = \pm \sqrt{\frac{\mathcal{E}^2}{r^2}\Upsilon(r)}\label{eq:drdtau}\,,
\end{align}
where we have defined
\begin{equation}
    \Upsilon(r)=\left(\frac{r^2-b^2A(r)}{A(r)B(r)}\right)\,.
\end{equation}
The ratio between Eq. \eqref{eq:dphidtau} and Eq. \eqref{eq:dtdtau} yields a differential equation directly relating the azimuthal coordinate, $\phi$, to the radial coordinate, $r$,
\begin{equation}
    \frac{d\phi}{dr} = \frac{b}{r\sqrt{\Upsilon(r)}}. 
    \label{eq:dphi_dr}
\end{equation}
Similarly, the ratio between Eq. \eqref{eq:drdtau} and Eq. \eqref{eq:dtdtau} yields a differential equation for the coordinate time, $t$, as a function of the radial coordinate, $r$,
\begin{equation}
    \frac{dt}{dr} = \frac{r}{A(r)\sqrt{\Upsilon(r)}}\,.
    \label{eq:dt_dr}
\end{equation}
The latter equation is the one that has to be integrated in order to compute the travel time, and thus the propagation delay, for a photon $\gamma$ originating at the emitter position and reaching the observer. 
In particular, given an emitter located at coordinates $E(r_{\rm e}, \phi_{\rm e})$ which emits a photon at coordinate time $t_{\rm e}$, and an observer located at $O(r_{\rm o}, \phi_{\rm o})$ which receives the photon at coordinate time $t_{\rm o}$ (see Fig. \ref{fig:emitter_observer}), we are interested in computing the travel time, $\Delta t = t_{\rm o}-t_{\rm e}$, taken by a photon to cover the distance from $E$ to $O$. Due to the curved photon path produced by the gravitational field of the central mass, one has to determine the appropriate value for the impact parameter $b$ of the photon $\gamma$ connecting points $E$ and $O$ in the curved space-time. This problem is known as the ``emitter-observer problem''. In order to approach this problem, one should have to integrate Eq. \eqref{eq:dphi_dr} over the trajectory of $\gamma$ and solve for $b$. This yields
\begin{equation}
    \phi_{\rm o}-\phi_{\rm e} = 
    \int_{r_{\rm e},\gamma}^{r_{\rm o}}\frac{b}{r\sqrt{\Upsilon(r)}}dr.
    \label{eq:phi_integral}
\end{equation}
The dependence of the integral in Eq. \eqref{eq:phi_integral} from the photon path $\gamma$ translates into the fact that one has to take into account whether or not the propagation of $\gamma$ from $E$ to $O$ is direct or indirect (see Fig. \ref{fig:emitter_observer} for the two different situations). This, in turn, depends on the specific geometrical configuration of the emitter and the observer with respect to the central object and, more precisely, on the fact that the radial component of the 4-velocity in Eq. \eqref{eq:drdtau} can be either positive or negative.
Considering only primary photons received by the observer (\textit{i.e.} we do not consider photons that graze so close to the unstable photon orbit of the central object that their paths bend so  strongly, $\Delta \phi > \pi$, that they reach the observer after one or more complete turns around the central object) only two possible scenarios are possible: the radial coordinate may increase monotonically going from $r_{\rm e}$ to $r_{\rm o}$ (we are assuming, as is usual, that the emitter is located closer to the central object than the observer), corresponding to a positive sign in Eq. \eqref{eq:drdtau}, in which case we will say that the propagation is \textit{direct}, thus resulting in
\begin{equation}
    \phi_{\rm o}-\phi_{\rm e} = \int_{r_{\rm e}}^{r_{\rm o}}\frac{b}{r\sqrt{\Upsilon(r)}}dr,
    \label{eq:phi_integral_direct}
\end{equation}
while, in the other case, the radial coordinate decreases (minus sign in Eq. \eqref{eq:drdtau}) up to a certain $r_{\rm min}$ (for which the right-hand side of Eq. \eqref{eq:drdtau} goes to zero) and then increases monotonically up to $r_{\rm o}$. In the latter case, the integral in Eq. \eqref{eq:phi_integral} would read
\begin{equation}
    \phi_{\rm o}-\phi_{\rm e} = \int_{r_{\rm e}}^{r_{\rm min}}\frac{b}{r\sqrt{\Upsilon(r)}} dr + \int_{r_{\rm min}}^{r_{\rm o}}\frac{b}{r\sqrt{\Upsilon(r)}}dr
    ,
    \label{eq:phi_integral_indirect}
\end{equation}
and we refer to this condition as \textit{indirect propagation}. As already mentioned in \citet{Hackmann2019}, it has been shown that a closed-form solution for such integrals exists for the case of a Schwarzschild BH that can be expressed in terms of Jacobi elliptic function (Eq. \ref{eq:exact_formula}). However, even if one solves 
the integrals analytically, solving the resulting equation for $b$ is not generally possible, and one has to resort to numerical methods 
to find the appropriate value of the impact parameter that solves the emitter-observer problem. Since our aim is to develop an algorithm that is able to compute the propagation time delay in any spherically symmetric space-time, we cannot even rely on the analytical solution of the integrals in Eq. \eqref{eq:phi_integral} that is valid for the Schwarzschild case, as $A(r)$ and $B(r)$, apart from satisfying some restriction to preserve the asymptotic flatness, could be functions of any sort. Moreover, in the Schwarzschild case, one can bring the right-hand side of Eq. \eqref{eq:drdtau} in a third-degree polynomial form \citep{Hackmann2019, Chandrasekhar1998}, thus reducing the problem of finding the radius $r_{\rm min}$ of closest approach for the indirect propagation case to that of finding polynomial-roots. In the general case, the function for which one has to find the root $r_{\rm min}$ can be indefinitely complicated and may not be provided at all with an analytic solution.\par
For these reasons, we approach the entire emitter-observer problem in a numerical fashion, building an algorithm that, given as inputs the coordinates for $E$ and $O$, does not only provide the impact parameter $b$ of the photon connecting the two points but is also able to autonomously determine whether it is the case of a direct or indirect propagation, giving, in the latter case, a numerically-computed value for $r_{\rm min}$. In order to do this, first of all, we consider the following: the radial coordinate $r_{\rm min}$ does not only represent a root for the right-hand side of Eq. \eqref{eq:drdtau}. Due to the fact that the radial coordinate decreases monotonically from $r_{\rm e}$ to $r_{\rm min}$ and it increases monotonically from $r_{\rm min}$ to $r_{\rm o}$ (up to $r_{\rm o}\to\infty$), it is also a local minimum for the radial coordinate. This guarantees that $r_{\rm min}$ is the largest among the roots of the right-hand side of Eq. \eqref{eq:drdtau} and the inequality $r_{\rm min} < b < r_{\rm e} < r_{\rm o}$ will always hold. Then, for a given value of $b$, we apply a root-finding algorithm, built upon the non-linear equation solver \textsc{Minpack} implemented in \citet{More1980}, using as an initial guess for $r_{\rm min}$ the value of $b$ itself, thus guaranteeing that the algorithm will converge to the root of the function that is the closest to $b$, and which thus corresponds to the value of $r_{\rm min}$ that we are seeking. Note that, in general, for values of $b$ below a critical impact parameter $b_{\rm crit}$ (which one usually relates to the apparent size of a BH shadow \citet{Chandrasekhar1998, EventHorizonTelescopeCollaboration2019,EventHorizonTelescopeCollaboration2022f}), the right-hand side of Eq. \eqref{eq:drdtau} does not have any real root in the range $r>0$. We check for this condition, and in case the algorithm converges to negative roots, we stop looking for indirect photons, as we are guaranteed that the propagation is direct.

Once the value of $r_{\rm min}$ is known, we can numerically compute the angular integrals in Eqs. \eqref{eq:phi_integral_direct} and \eqref{eq:phi_integral_indirect}, that, for generic radial coordinates $r_1$ and $r_2$ and a given value of $b$ read as
\begin{equation}
    \Phi(r_1, r_2; b)  =\int_{r_1}^{r_2}\frac{b}{r\sqrt{\Upsilon(r)}}dr\,,
    \label{eq:Phi_integral}
\end{equation}
by employing the \textsc{Quadpack} routine implemented in \citet{Piessens1983}. In particular, we do this for both the direct photon and the indirect one (provided that the latter exists, \textit{i.e.} that the specific value of $b$ is above its critical value) and compute
\begin{align}
    &\Delta\phi_{\rm direct}(r_{\rm e}, r_{\rm o}; b) = \Phi(r_{\rm e}, r_{\rm o}; b),\\
    &\Delta\phi_{\rm indirect}(r_{\rm e}, r_{\rm o}; b) = \Phi(r_{\rm e}, r_{\rm min}; b) + \Phi(r_ {\rm min}, r_{\rm o}; b).
\end{align}
We thus have a routine that for fixed values of $r_{\rm e}$ and $r_{\rm o}$ and a given value of $b$ provides us with two numerical estimations of the angular distance between emitter and observer. We can apply again the \textsc{Minpack} root-finder routine \citep{More1980} to look for the value of $b$ that solves the equation
\begin{equation}
    \Delta\phi-(\phi_{\rm o}-\phi_{\rm e}) = 0\,,
\end{equation}
where, at each iteration, $\Delta \phi$ is the one between $\Delta\phi_{\rm direct}$ and $\Delta\phi_{\rm indirect}$ that is closer to the target solution $(\phi_{\rm o}-\phi_{\rm e})$. Here, we use as an initial guess for the impact parameter $b_0 = r_{\rm e}\sin(\phi_{\rm o}-\phi_{\rm e})$, which corresponds to the impact parameter in flat space-time. Clearly, the choice between $\Delta\phi_{\rm direct}$ and $\Delta\phi_{\rm indirect}$ for the last iteration of the root-solver will determine the direct/indirect nature of the resulting photon path. 

\begin{figure}
    \centering
    \includegraphics[width = \columnwidth]{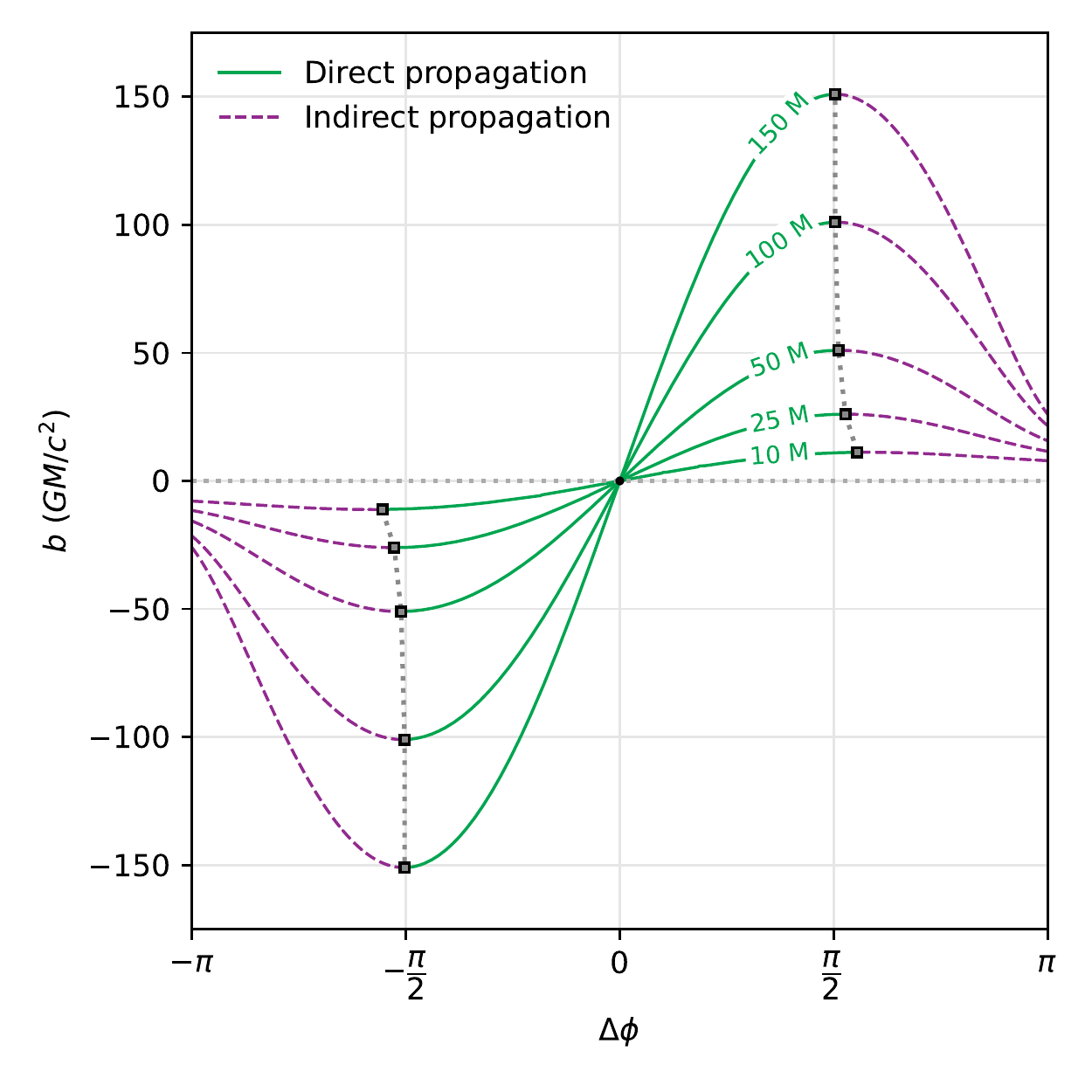}
    \caption{Example of use of our routine for solving the emitter-observer problem in a Schwarzschild space-time with mass $M$. For a fixed observer (located at $r_{\rm o}=10^9M$), we have considered different radial coordinates of the emitter $r_{\rm e}$ between 10 $M$ and 150 $M$ (reported in green on the corresponding lines) and for values of $(\phi_{\rm o}-\phi_{\rm e})$ in the range $[0,2\pi]$. In particular, green solid lines correspond to photons that propagate directly from $E$ to $O$, and violet dashed lines to photons that undergo indirect propagation. For each value of $r_{\rm e}$, we have highlighted (with gray squares) the separation between direct and indirect propagation. This point of separation tends towards $\Delta\phi = \pi/2$ as the distance of the emitter from the central object increases (meaning that for a sufficiently far emitter, the entire semi-plane containing the observer corresponds to a direct propagation) and deviates  from $\pi/2$ as the emitter gets close to the center, due to the strong bending of photon paths.}
    \label{fig:emitter_observer_b_phi}
\end{figure}

Once the emitter-observer problem has been solved, we have access to the impact parameter $b$ of the photon corresponding to the primary image of the source for the observer in $O$. We can now approach the problem of determining the travel time of the photon from $E$ to $O$. Here the assumption of asymptotic flatness of the space-time in Eq. \eqref{eq:generic_metric} is crucial for two reasons. First, it allows us to assume that for a sufficiently far observer (\textit{i.e.} for an Earth-based observer, the distance from SgrA* is $\sim8\textrm{ kpc} = 4\times10^{10}$ 
gravitational radii of the source) the curvature of space-time produced by the compact central object can be assumed to be zero at the observer location. Of course, in the case of the Earth, one should take into account the presence of the Sun's gravitational field and the motion of Earth around it. However, this contribution can be added later using classical formulas (\textit{e.g.} from \cite{Damour1986}) since we can consider a weak field approximation for the Sun. As a consequence of this, we can assume that the observer actually measures the coordinate time $t$ and thus the travel time is simply the integral of Eq. \eqref{eq:dt_dr} over the photon's path $\gamma$:
\begin{equation}
    \Delta t \equiv t_{\rm o}-t_{\rm e} = \int_{r_{\rm e},\gamma}^{r_{\rm o}} \frac{r}{A(r)\sqrt{\Upsilon(r)}}dr\,,
    .
\end{equation}
Again, as done for the integrals of the angular coordinate, the information on the photon path $\gamma$ is encoded in the impact parameter $b$ that corresponds to the one resulting from the solution of the emitter-observer problem and in the fact that one either integrates directly from $r_{\rm e}\to r_{\rm o}$ in the direct propagation case, or passing by $r_{\rm int}$ (\textit{i.e.} over the radial path $r_{\rm e}\to r_{\rm int} \to r_{\rm o}$) in the indirect one. Considering the solution of the integral between two generic radial coordinates:
\begin{equation}
    T(r_1, r_2) =  \int_{r_1}^{r_2} \frac{r}{A(r)\sqrt{\Upsilon(r)}}dr\,,
    \label{eq:T_integral}
\end{equation}
we can express the two cases by
\begin{align}
    &\Delta t_{\rm direct} = T(r_{\rm e}, r_{\rm o}),\\
    &\Delta t_{\rm indirect} = T(r_{\rm e}, r_{\rm min})+T(r_{\rm min}, r_{\rm o}).
\end{align}
As for the case of the emitter observer problem, while an analytic solution in terms of Jacobi elliptic function exists for the case of the Schwarzschild space-time \citep{Hackmann2019}, in the most general case, one has to approach the problem numerically. However, differently from the previous integral, the integration of Eq. \eqref{eq:T_integral} presents additional challenges from a computational perspective. The integrand function in Eq. \eqref{eq:Phi_integral} tends to $0$ when $r\to\infty$, thus the numerical quadrature method implemented in \textsc{Quadpack} is
always able to converge within the desired precision tolerance. The integrand of $T$ in Eq. \eqref{eq:T_integral}, on the other hand, generally tends to a non-null constant value for $r\to\infty$. This is, from a physical point of view, related to the fact that since this integral returns the travel time for a photon, as $r_{\rm o}$ grows, the travel time has to grow accordingly due to the additional time required to cover the extra distance. This condition, however, can result in a misbehavior of the \textsc{Quadpack} routine leading to round-off errors, especially when the range of integration is particularly large (as is the case for a \textit{very distant} observer). Here, again, the asymptotic-flatness assumption can help us sort this issue out. As a matter of fact, regardless of the specific functional form of $A(r)$ and $B(r)$, in an asymptotically-flat space-time both of these functions approach $1$ as $r\to\infty$ \citep{Wald1984}, and subsequently the integrand itself of Eq. \eqref{eq:T_integral} tends to $1$. We can thus express the integral in the following way
\begin{equation}
    T(r_1, r_2) = \int_{r_1}^{r_2}\left(\frac{r}{A(r)\sqrt{\Upsilon(r)}}-1\right)dr + (r_2-r_1),
    \label{eq:T_integral_as_flat}
\end{equation}
and, taking advantage of the fact that the integrand now tends to $0$ for $r\to\infty$, we can integrate the first term numerically within any desired precision tolerance.

\subsection{Time delay in a Schwarzschild space-time}

\begin{figure}
    \centering
    \includegraphics[width = \columnwidth]{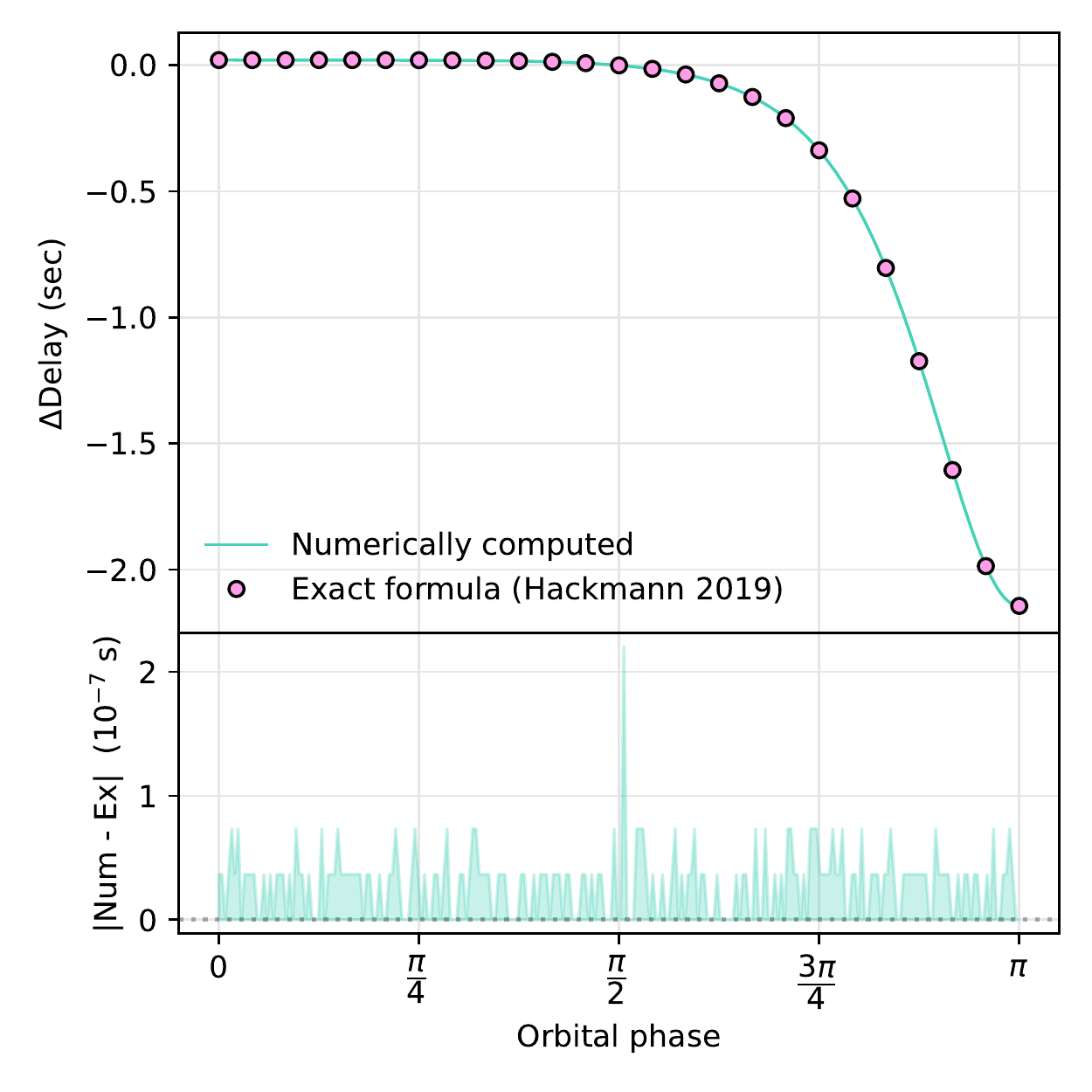}
    \caption{\textit{Top panel:} difference in seconds between the fully relativistic delay and the post-Newtonian approximation for both the exact formula (pink dots) and our numerical routine (aqua solid line) as a function of the orbital phase for the model Toy 0 from \citet{Hackmann2019}. The plot shows perfect agreement between the two predictions. \textit{Bottom panel:} difference in units of $10^{-7}$ s between our numerical estimation of the photon propagation time and that by applying exact formula by \citet{Hackmann2019}.}
    \label{fig:exact_vs_numerical}
\end{figure}

In order to validate our methodology, we compute the  time delay for a specific emitting object around a 
supermassive Schwarzschild BH using our approach and compare 
the results to those obtained with the exact analytic formula by \citet{Hackmann2019} in Eq. \eqref{eq:exact_formula}. 
We particularize our calculations to Sgr A* and, therefore, we set the mass to $M = 4\times 10^6M_\odot$.
Firstly, in Fig. \ref{fig:emitter_observer_b_phi}, we report an example of usage of our routine for solving the emitter-observer problem for the Schwarzschild case (nonetheless, due to the generality of the approach we adopted, the routine works with any choice for $A(r)$ and $B(r)$). The observer $O$ is located at $r_{\rm o}=10^9M$ (we are setting $G=c=1$ for convenience) and $\phi_{\rm o} = 0$ and we have considered different radial coordinates of the emitter $r_{\rm e}$ between 10 $M$ and 150 $M$ and values of $\Delta\phi = \phi_{\rm o}-\phi_{\rm e}$ in the range $[0,2\pi]$.  
We highlight the angular distance for each radial distance for which photons propagate directly from $E$ to $O$ (green solid lines) and photons that undergo indirect propagation (purple dashed lines). Moreover, the angular distance corresponding to the separation between direct and indirect propagation is reported. This tends towards $\Delta\phi = \pi/2$ as the distance of the emitter from the central object increases: for a distant emitter, photons propagate directly to the observer when starting in the entire semi-plane containing the observer itself. For an emitter that is closer to the central object, this separation deviates from $\pi/2$ due to the strong lensing effects that bends photon paths. 
Next, we consider the same circular orbit with $r = 100 M$ 
and inclination $i = \pi/3$ considered in \citet{Hackmann2019} (we will refer to the such orbiting body as Toy 0 for the rest of this work). 
All orbital parameters are summarized in Table \ref{tab:toy_models}. 
In this case, the object's trajectory is considered a Keplerian circular orbit on which no relativistic effect
(such as the orbital precession) is taken into account (the compact central SMBH only affects the propagation of photons). We can thus parametrize positions on the orbit using as orbital phase the mean anomaly
\begin{equation}
    \varphi = \frac{2\pi}{T}(t-t_p),
    \label{eq:orbital_phase}
\end{equation}
where $T$ is the orbital period (derived by applying Kepler's third law), and $t_p$ is the time of pericenter passage that we set for convenience as $t_p = 0$. We considered half of the orbit $\varphi\in[0,\pi]$ and computed the corresponding time of travel for photons to reach an Earth-based observer located at $r_{\rm o} = 8$ kpc from the central object, both using our numerical routine and with the exact formula. Then, following \citet{Hackmann2019}, we subtract to the obtained quantities the classical Röemer time of travel and the Shapiro delay using the usual post-Newtonian expressions \citep{Damour1986} and, choosing the point on the orbit corresponding to an orbital phase of $\phi = \pi/2$ as a reference point, we subtract the corresponding time of travel. The resulting difference between the fully relativistic delay and the post-Newtonian approximation is shown in Fig. \ref{fig:exact_vs_numerical} for both the exact formula and our numerical routine. 
This difference estimates the error committed when applying a post-Newtonian approximation to the photon propagation time and, thus, would correspond to the timing residuals of a post-Newtonian fit to possible GC pulsar data.
The numerical and exact profiles perfectly agree with each other and with the one shown in Fig. 1 of \citet{Hackmann2019} as also confirmed by the very small residuals ($\lesssim 10^{-7}$ s) reported in the bottom panel consistent with the numerical tolerance of the algorithm, thus validating our methodology for the Schwarzschild case.

\section{A brief overview on alternatives to a Schwarzschild black hole space-time}
\label{sec:alternatives}

After checking 
that our pipeline for the computation of the propagation time delay 
works as expected in the Schwarzschild space-time, we moved to apply the developed methodology to a series of spherically symmetric space-times that represent alternatives to the standard BH paradigm. In particular, we have chosen to analyze the behavior of three specific space-times: the Black Bounce (BB, Sec. \ref{sec:BB}) that, formulated within GR, considers possible alternative natures for the central object; the BH solution in Scalar-Tensor-Vector Gravity (STVG, Sec. \ref{sec:STVG}), an extended theory of gravity where additional scalar and vectorial degrees of freedom are taken into account; and a BH solution in Einstein-Maxwell-dilaton-axion (EMDA, \ref{sec:EMDA}), that is an alternative theory of gravity arising from a string theory scenario.

\subsection{The Black Bounce space-time}
\label{sec:BB}
One possible alternative to the standard BH paradigm, within the general relativistic framework, is the BB model. Firstly introduced in \citet{Simpson2019}, this one-parameter family of solutions is described by the following line element:
\begin{equation}
    ds^2 = -\left(1-\frac{2M}{\sqrt{x^2+\alpha^2}}\right)dt^2+\left(1-\frac{2M}{\sqrt{x^2+\alpha^2}}\right)^{-1}dx^2+(x^2+\alpha^2)d\Omega^2,
    \label{eq:BB_metric}
\end{equation}
where we have used geometric units $G_N = c = 1$ and $d\Omega ^2 = d\theta^2+\sin^2\theta d\phi^2$ 
is the solid angle element. The space-time coordinates $(t, x, \theta, \phi)$ in Eq. \eqref{eq:BB_metric}, are defined in the following intervals
\begin{align}
    &t\in(-\infty,+\infty),& &x\in(-\infty,+\infty),& &\theta\in(0,\pi),& &\phi\in(0,2\pi),
\end{align}
and, considering the line element $ds^2_{(2)}$ on an hypersurface characterized by $\{t= \textrm{const.}; x = \textrm{const.}\}$,  $(\theta,\phi)$ define usual spherical angular coordinates on a sphere with area $A = \int ds^2_{(2)} = 4\pi r^2$, where $r^2(x) = x^2 + \alpha^2$ can be regarded as the aerial radius in these coordinates. The extra parameter $\alpha$ defines the amplitude of the throat of the geometry, $r_{\rm th}^2 = \alpha^2$, $\textit{i.e.}$ the absolute minimum of the aerial radius. Interestingly, the presence of this bouncing feature in the geometry, whose properties vary with the value of $\alpha$, allows the static and spherically symmetric family of solutions in Eq. \eqref{eq:BB_metric} to describe objects of different nature \citep{Simpson2019}. Indeed, setting $\alpha=0$ returns the Schwarzschild BH metric. However, for values of $\alpha\neq 0$, the metric smoothly interpolates between a Schwarzschild BH and a traversable 
WH. To summarize:
\begin{itemize}
    \item When $\alpha = 0$, $x = r$ and the geometry has a horizon in $r_H = \pm 2M$ (given by the roots of the $g_{00}$ coefficient), thus reducing to the usual Schwarzschild BH with mass $M$;
    \item When $0 < \alpha < 2M$, in the metric a horizon and a non-null throat appears, however $|r_H| > |r_{\rm th}|$ and so that the bounce is \textit{hidden} by the horizon. Moreover, for $x\to0$ the metric coefficients do not diverge and, hence, the solution corresponds to a regular BH geometry \citep{Roman1983, Hayward2006};
    \item When $\alpha = 2M$, we have $|r_H| = |r_{\rm th}|$, so that the two regions $x<0$ and $x>0$ are connected at the throat (and thus the geometry represents a WH) but since the surface $x = 0$ is null (\textit{i.e.} a horizon) it is only one-way traversable \citep{Cano2019, Simpson2019};
    \item Finally, when $\alpha > 2M$, the geometry is horizonless, 
    and the bounce is exposed on both sides. The geometry thus represents a two-way traversable WH \citep{Morris1988a, Morris1988b, Visser1989a, Visser1989b}. 
\end{itemize}
The value $\alpha = 2M$, hence, represents the separation between the BH/WH classes of geometries in this family. The metric in Eq. \eqref{eq:BB_metric} has been tested at the GC by studying the geodetic motion of the S-stars orbiting Sgr A*. \citet{DellaMonica2022b} showed that currently publicly available data for the S2 star in the GC are only able to place an upper limit on the parameter $\alpha$ of order $\alpha \lesssim 140 M$ at 95\% confidence level, thus not being able to unambiguously recognize the nature of the central object. Remarkably, a forecast analysis within the same study demonstrates that much more precise data for the S2 star from the GRAVITY interferometer are expected to improve by a factor $\sim 25$ this upper limit (still being unable to distinguish a BH from a WH) and that only studying the orbits of putative much closer stars would allow saying something conclusive about the WH/BH nature of Sgr A*. 
Interestingly, in \citet{Guerrero2021}, light rings and shadows resulting from the BB model have been studied, highlighting some differences in the optical appearance of the BB solutions as compared to the Schwarzschild one.

Here, we wish to investigate  the sensibility of the propagation time delay of a pulsar orbiting Sgr A* on the parameter $\alpha$, for which we probe an interval $\alpha\in [0,4]M$ centered on the value of separation between a BH and a two-way traversable WH, $\alpha = 2M$, and use the methodology depicted in Sec. \ref{sec:alternatives}. It is worth mentioning that the metric that appears in Eq. \eqref{eq:BB_metric} is not directly expressed in the form of our generic spherically symmetric model in Eq. \eqref{eq:generic_metric}. However, by considering the definition of the aerial radius $r^2 = x^2+\alpha^ 2$ it is easy to bring the BB metric in the desired form, resulting in:
\begin{align}
    A(r) &= \left(1-\frac{2M}{r}\right),\\
    B(r) &= \frac{r^2}{A(r)(r^2-\alpha^2)}.
\end{align}

\subsection{The Scalar-Vector-Tensor Gravity}
\label{sec:STVG}

Considering extra scalar, tensorial, and vectorial degrees of freedom in the GR action is a viable way of formulating meaningful extensions to GR \citep{deLaurentis2023}. This is the case, for example, of STVG. As its name suggests, this modified theory of gravity, first presented in \citet{Moffat2006}, considers the introduction of extra degrees of freedom in the description of the gravitational interaction in the form of scalar fields and a massive vector field $\phi^\mu$ (along with metric tensor field $g_{\mu\nu}$). In particular, while the gravitational constant $G$ and the mass $\mu$ of the vector field are elevated to the role of scalar fields, a vector field $\phi^\mu$ that couples with massive test particles (of mass $m$) are introduced to encode a fifth-force interaction that modifies the geodesic equation
\begin{align}
    \left(\frac{d^2x^\mu}{d\lambda^2}+\Gamma^\mu_{\nu\rho}\frac{dx^\nu}{d\lambda}\frac{dx^\rho}{d\lambda} \right)=\frac{q}{m}{B^{\mu}}_\nu\frac{dx^\nu}{d\lambda}.
    \label{eq:geodesic-equations_mog}
\end{align}
Here, $q$ is the charge by which matter couples to the vector field $\phi^\mu$, encoded in the tensor $B_{\mu\nu}:= \nabla_\mu\phi_\nu-\nabla_\nu \phi_\mu $. In STVG, the generally covariant action can be written as \citep{Moffat2006}
\begin{equation}
    \mathcal{S} = \mathcal{S}_{GR}+\mathcal{S}_M+\mathcal{S}_\phi+\mathcal{S}_S.
    \label{eq:stvg-action}
\end{equation}
The first two terms are the classical Hilbert-Einstein action of GR and the action for ordinary matter. The two additional terms, $\mathcal{S}_\phi$ and $\mathcal{S}_S$, encode the novel features of STVG
\begin{align}
    \mathcal{S}_\phi =& -\int d^4x\sqrt{-g}\left(\frac{1}{4}B^{\mu\nu}B_{\mu\nu}-\frac{1}{2}\mu^2\phi^\nu\phi_\nu+V(\phi)\right),\\
    \mathcal{S}_S =& \int d^4x\sqrt{-g}\frac{\omega_M}{G^3}\left(\frac{1}{2}g^{\mu\nu}\nabla_\mu G\nabla_\nu G-V(G)\right)+\nonumber\\
    &+\int d^4x\frac{1}{\mu^2G}\left(\frac{1}{2}g^{\rho\nu}\nabla_\rho\mu\nabla_\nu\mu-V(\mu) \right).
\end{align}
Here, $g$ stands for the determinant of the metric tensor $g_{\mu\nu}$, $\omega_M$ is a constant, and $V(\phi)$, $V(G)$ and $V(\mu)$ are scalar potentials arising from the self-interaction associated with the vector field and the scalar fields, respectively.  From the minimization 
of the Eq. \eqref{eq:stvg-action}, the field equations in vacuum ({\em i.e.} $T_{\alpha\beta}^M = 0$) read \citep{Moffat2021}
\begin{align}
    G_{\mu\nu}=&-\frac{\omega_M}{\chi^2}\biggl(\nabla_\mu\chi\nabla_\nu\chi -\frac{1}{2}g_{\mu\nu}\nabla^\sigma\chi\nabla_\sigma\chi\biggr)+
\\&-\frac{1}{\chi}(\nabla_\mu\chi\nabla_\nu\chi-g_{\mu\nu}\Box\chi)+\frac{8\pi}{\chi}T^\phi_{\mu\nu},
\end{align}
where the scalar field $\chi = 1/G$, and $T^\phi_{\alpha\beta}$ is the gravitational $\phi$-field energy momentum tensor given by
\begin{equation}
    T^\phi_{\mu\nu} = -\left({B_\mu}^\sigma B_{\sigma\nu}-\frac{1}{4}g_{\mu\nu}B^{\sigma\rho}B_{\sigma\rho}\right).
\end{equation}

\citet{Moffat2015} derived an explicit BH solution in STVG. Among the hypotheses that are involved in such derivation, $G$ can be regarded as a constant with a larger value than its Newtonian counterpart, $G_N$, encoded in an additional dimensionless parameter $\alpha$, $G = G_{\rm N}(1+\alpha)$. Moreover, it is postulated that the mass $\mu$ of the vector field can be neglected on the scales of compact objects and that the fifth-force charge $q$ of a test particle can be written as $q = m\sqrt{\alpha G_{\rm N}}$. From these assumptions, a spherically symmetric space-time metric is derived,
\begin{equation}
    ds^2 = -\frac{\Delta}{r^2}dt^2+\frac{r^2}{\Delta}dr^2+r^2d\Omega^2,
\end{equation}
with $\Delta = r^2-2Mr+\alpha M\left((1+\alpha)M-2r\right)$ and $d\Omega^2 = d\theta^2 + \sin^2\theta d\phi^2$. This space-time is formally equivalent to that of a Reissner–Nordström charged BH \citep{Reissner1916} where in this case, the charge is the fifth-force charge of the central object and reduces to the usual Schwarzschild metric when $\alpha = 0$. It is straightforward to put the metric in our notation:
\begin{align}
    A(r) &= \left(1-\frac{2M}{r}+\frac{\alpha M}{r^2}((1+\alpha)M-2r)\right),&
    B(r) &= \frac{1}{A(r)}.
\end{align}
When studying the free-fall motion of massive (and thus fifth-force-charged) particles in STVG and the interaction with the vector field $\varphi^\mu$, it is possible to derive a first-order analytical expression for the rate of orbital precession, that is given by \cite{DellaMonica2023b}
\begin{equation}
    \Delta\omega_{\rm STVG} = \Delta\omega_{\rm GR}\left(1+\frac{5}{6}\alpha\right).
    \label{eq:precession_STVG}
\end{equation}
The precession scales linearly with the parameter $\alpha$ and reduces to the usual GR expression in Equation \eqref{eq:gr_precession} when $\alpha = 0$. Studying the geodesic motion of the S2 star in the GC around Sgr A* and taking advantage of the measurement of the orbital precession measured in \citet{GRAVITYCollaboration2020}, in \citet{DellaMonica2022a} a stringent upper limit for the parameter $\alpha$ is derived, $\alpha \lesssim 0.662$ at 99.7\% confidence level, thus providing the first constraints for STVG on such scales. Here, we want to show that the timing analyses of putative pulsars around the same object would be able to improve significantly such constraints  and, for this reason, we investigate the propagation delay around an STVG SMBH in the range $\alpha\in[0,0.1]$.

\subsection{The Einstein-Maxwell-dilaton-axion theory of gravity}
\label{sec:EMDA}
The EMDA theory arises from the low-energy Lagrangian of the superstring theory \citep{Sen1992,GarciaGaltsov1995,Tripathi2021}.
Specifically, it arises as the resulting bosonic part of compactifying the ten-dimensional heterotic string theory on a six-dimensional torus. The theory includes a pseudo-scalar axion field and a scalar dilaton field coupled to the Maxwell field and the metric, which leads to observational implications. For instance, the deflection of light by the gravity of a BH is qualitatively different between EMDA and standard Kerr BHs, leading to measurable differences in their gravitational lensing observables \citep{Gyulchev2007, Mizuno2018}.
The action $\mathcal{S}$, associated with EMDA, gravity contains couplings of the metric $g_{\nu\mu}$, the U(1) gauge field $A_\mu$, the dilaton field $\chi$, and an anti-symmetric tensor field $H_{\rho\sigma\delta}$ related to the axion pseudo-scalar field $\xi$. In four dimensions, the action can be expressed as
\begin{align}
\mathcal{S}=\frac{1}{16\pi}\int \sqrt{-g}d^4x \biggl(&R-2\partial_{\nu}\chi\partial^{\nu}\chi-\frac{1}{2}e^{4\chi}\partial_{\nu}\xi\partial^{\nu}\xi+\nonumber\\
&+e^{-2\chi}F_{\rho\sigma}F^{\rho\sigma}+\xi F_{\rho\sigma}\tilde{F}^{\rho\sigma} \biggr),
\end{align}
where $g$ is the determinant of the metric, $R$ is the Ricci scalar with respect to the metric, and $F_{\mu\nu}$ is the Maxwell field strength tensor defined as $F_{\mu\nu}=\nabla_{\mu}A_{\nu}-\nabla_{\nu}A_{\mu}$.
The variation of the action with respect to the metric gives rise to the field equations in this theory, which are of the form
\begin{equation}
G_{\mu\nu}=T_{\mu\nu}(F,\chi,\xi),,
\end{equation}
where $G_{\mu\nu}$ defines the Einstein tensor, and the energy-momentum tensor $T_{\mu\nu}$ is given by:
\begin{align}
T_{\mu\nu}(F,\chi,\xi)=&e^{2\chi} (4F_{\mu\rho}F^{\rho}{\nu}-g{\mu\nu}F^2 )+\nonumber\\
&-g_{\mu\nu} ( 2 \partial_\gamma\chi\partial^\gamma\chi+\frac{1}{2}e^{4\chi}\partial_\gamma\xi\partial^\gamma\xi)+\nonumber\\
&+\partial_\mu\chi\partial\nu\chi+e^{4\chi}\partial\mu\xi\partial_\nu\xi.
\end{align}

\citet{Sen1992} derived a stationary and axisymmetric solution of the field equations also known as the Kerr-Sen metric which, in the Boyer-Lindquist coordinate system, is given by
\begin{align}
ds^2 = & -\left( 1-\frac{2M\tilde{r}}{\tilde{\Sigma}} \right)dt^2 +\frac{\tilde{\Sigma}}{\Delta}(d\tilde{r}^2+\Delta d{\theta}^2)-\frac{4aM\tilde{r}}{\tilde{\Sigma}}\sin^2\theta dtd\phi+\nonumber\\
&+\sin^2\theta d\phi^2 \left( \tilde{r}(\tilde{r}+r_2)+a^2+\frac{2M\tilde{r}a^2\sin^2\theta}{\tilde{\Sigma}} \right),
\end{align}
where
\begin{align}
& \tilde{\Sigma}=\tilde{r}(\tilde{r}+r_2)+a^2\cos^2\theta, \\
& \Delta=\tilde{r}(\tilde{r}+r_2)-2M\tilde{r}+a^2.
\end{align}
Here, $M$ represents the mass, $r_2=\frac{q^2}{M}e^{2\chi_0}$ is the dilaton parameter, and $a$ is the dimensionless spin parameter of the BH. The dilaton parameter contains information about both the asymptotic value of the dilaton field $\chi_0$ and the electric charge $q$ of the BH, which arises from the coupling of the photon with the axion pseudo-scalar. When $q=0$, the Kerr metric is recovered. It can be shown that BH solutions must satisfy the following restriction on the parameter $r_2$ \citep{Banerjee2021}:
\begin{equation}
0\leq  {r_2} \leq 2M,
\end{equation}
where the value $r_2=0$ would recover GR.
Since we are interested in the propagation delay in a spherically symmetric space-time, we consider a Kerr-Sen BH without rotation, where both the spin and the axionic field vanish. In this way, a pure dilaton BH is obtained, whose deviation from Schwarzschild is encoded into the dilaton parameter which, as shown in \citet{Mizuno2018}, can be rewritten as $b={r_2}/{2}$ in the spinless BH limit (theoretically bound to $0\leq b \leq M$).  Thus, the space-time metric takes the form \cite{Mizuno2018}:
\begin{equation}
ds^2=- \left(\frac{\tilde{r}-2\mu}{\tilde{r}+2b}\right)dt^2
+ \left(\frac{\tilde{r}+2b}{\tilde{r}-2\mu} \right)dr^2
+ (\tilde{r}^2+2b\tilde{r})d\Omega^2,
\end{equation}
where $d\Omega^2$ is the solid angle, and the pseudo-radial coordinate $r$ and mass $M$ are defined as:
\begin{align}
& r^2=\tilde{r}^2 + 2b\tilde{r}, \
& M=\mu+b.
\end{align}
Therefore, to bring the EMDA metric in the form of Eq. \eqref{eq:generic_metric}, we must define
\begin{align}
    A(r) &= 1-\frac{2M}{b+\sqrt{b^2+r^2}},&
    B(r) &= \frac{1}{A(r)},
\end{align}
and thus, apart from the mass $M$ it only depends on the dilatonic parameter $b$, which we vary in its theoretical allowed range of $b\in[0,1]M$ (more details about the specific ranges of values used are reported in Sec. \ref{sec:results}).

\section{Results}
\label{sec:results}

\begin{figure*}
    \centering
    \includegraphics[width = \textwidth]{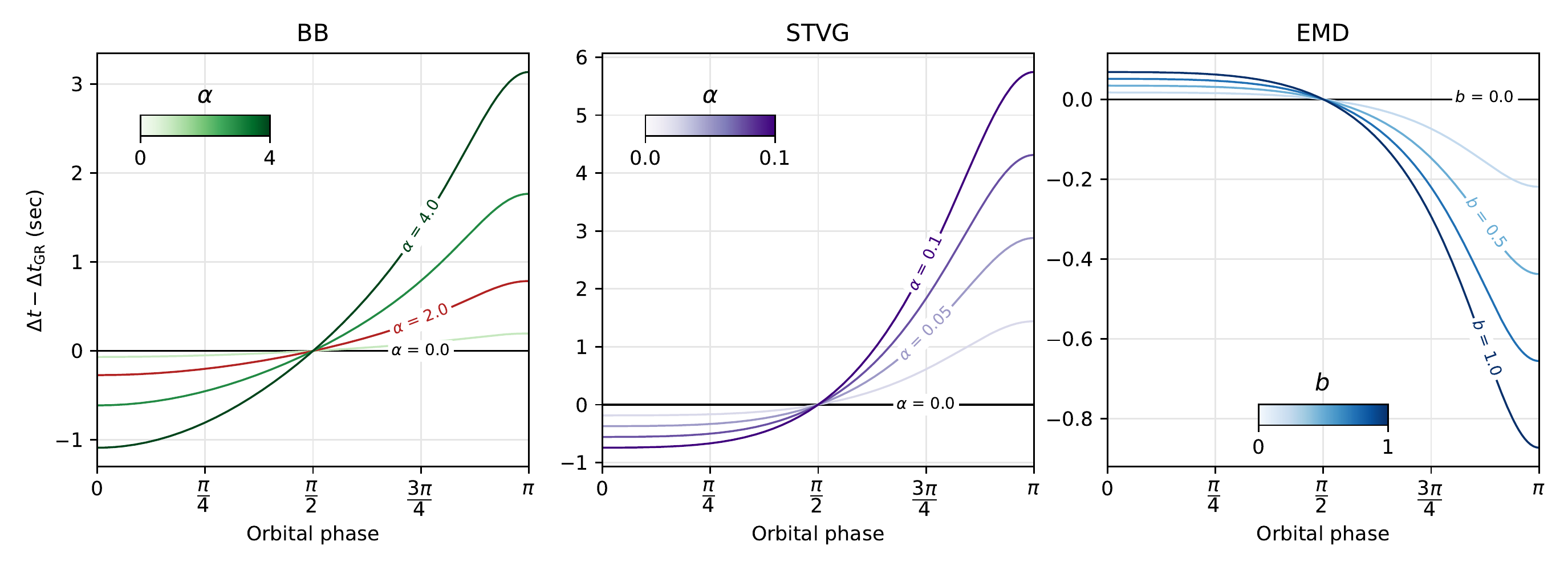}
    \caption{Differences in seconds from the Schwarzschild case of the propagation delay in the BB (green lines, left panel. The red line corresponding to $\alpha = 2M$ marks the separation between BH and WH solutions in the model), STVG (purple lines, center panel), and EMD (blue lines, right panel) models, respectively, computed numerically with our code for a central object of $M=4\times 10^6M_\odot$ and for the orbiting object Toy 0. The 0-level black solid lines correspond to the GR/Schwarzschild limit of each model and correspond to 
    the same propagation delay reported in Fig. \ref{fig:exact_vs_numerical} from \citet{Hackmann2019}. Colored solid lines correspond to increasingly higher deviations from the Schwarzschild limit obtained for different values of the theory parameter in their range of interest, as reported in the color bars and on the plotted labels.}
    \label{fig:delays_vs_pn}
\end{figure*}

We computed the photon propagation time for the BB, STVG, and EMDA models in order to assess differences from the Schwarzschild case and to check for the appearance of peculiar features that arise in the propagation delay for each model. 
To do so, 
as a first step, we particularized our procedure to the Toy 0 object reported in Table \ref{tab:toy_models}, and computed the difference between the resulting propagation delay and the Schwarzschild case. 
The results of our calculations are reported in Fig. \ref{fig:delays_vs_pn}, for the three models and for values of the extra parameters in their respective ranges. In particular, for all models we subtract the propagation delay in the Schwarzschild case (obtained in all cases by making the theory parameter approaches zero) from that obtained in correspondence with four values of the theory parameter, uniformly sampling the range of interest. 
Without loss of generality, we subtract a constant value in order to reduce to zero the difference between the two values at $\varphi = \pi/2$ as in \citet{Hackmann2019}. 
In all the cases, a higher value of the theory parameter corresponds to a bigger deviation from the Schwarzschild case. Moreover, the greatest difference between the computed delay and the Schwarzschild case is achieved for $\varphi = \pi$, \textit{i.e.} when the pulsar is farther away from the observer and, hence, the photon has to travel a maximal path (\textit{i.e.}, the longest indirect possible path between emitter and observer) in the SMBH gravitational field (from which the higher deviation arises). These 
results show that, even for an orbit like that of Toy 0, which is perfectly circular and thus not exhibiting relativistic effects on the orbit itself, the propagation delay does carry significant information about the space-time metric that can lead to differences of the order of up to a few seconds in the photon arrival time. For example, in the BB model, this difference corresponds to $\sim 1$ s for $\alpha = 2M$ (red line) and can reach $\sim 3$ s for $\alpha = 4M$, thus providing a measurable effect related to the nature of the central object.

While the results obtained for Toy 0 show that it is indeed possible to detect signatures of a modification of space-time at a metric level on the propagation delay of photons emitted by a pulsar, such a system might not be the ideal tool to investigate these effects. First, it does not possess realistic orbital features that are expected to belong to these types of objects in the GC. In fact, known stellar populations in the GC that may serve as progenitors for a pulsar population \citep{Eatough2015} exhibit especially high eccentricities \citep{Gillessen2009} that are thus expected to persist in the pulsars therein. 
Consequently, a pulsar on a highly eccentric orbit would experience a much broader class of relativistic effects with respect to our Toy 0. Indeed, the fact that this putative object keeps the same distance from the central object along its circular orbit makes it so that the impact of the relativistic orbital effect of periastron advance (Eq. \ref{eq:gr_precession}) and Einstein delay (Eq. \ref{eq:einstein_delay}) amounts to a constant shift between the pulsar's proper time and the coordinate time, that is totally degenerate with the intrinsic period of the pulses. For an object on a highly eccentric orbit, on the other hand, the orbital precession has a significant contribution and, as it represents a secular effect, the amplitude of the deviations from a Keplerian orbit grows over time \citep{Will2014}. On the contrary, the special and general relativistic time dilation effects have the same amplitude on each orbital period but do vary along the orbit due to the fact that the distance from the central object is not fixed, resulting in a phase-dependent additional delay. Moreover, since such effects are directly related to the metric coefficients, they can result in significant signatures from the underlying space-time geometry on the timing profiles. To investigate this possibility, we considered three different orbital models for pulsars in the GC, first introduced in \citet{DeLaurentis2018b}, whose orbital parameters are reported in Table \ref{tab:toy_models}. These are meant to explore an increasingly strong field regime for gravity by reaching increasingly closer pericenter distances and, thus, a deeper dive into the gravitational field of the central object. For these models, we have considered three different orbital inclinations of $i=0^\circ$ (edge-on configuration), $30^\circ$, and $60^\circ$ in order to explore how different geometrical configurations between emitter and observer alter the resulting time delay. For each configuration, we have numerically integrated the geodesic equation in Eq. \eqref{eq:geodesic_eq} (and Eq. \eqref{eq:geodesic-equations_mog} in the case of STVG) for the trajectory of the pulsar, in order to take into account for all the relativistic effects on the pulsar orbit and on its pulse emission times. The procedure followed for the numerical integration is the same adopted in previous works \citep{DeMartino2021, DellaMonica2022a, DellaMonica2022b}: we assign initial conditions at a given time (that we assume to correspond to the apocenter passage) starting from the Keplerian orbital elements of the pulsar, which corresponds to the Keplerian ellipse that osculates the true relativistic trajectory at the initial time. We carry out the numerical integration of the geodesic for several orbital periods for each model, and then, for each position on the orbit of the pulsar, we have applied the methodology developed in Sec. \ref{sec:alternatives} to compute the corresponding photon propagation delay. Due to the prograde periastron advance, the time $T$ it takes for the pulsar to make a full revolution around the central SMBH (\textit{i.e.} the time required to span an angle $2\pi$ on the orbital plane) is less than the time $T_{\rm pre}$ it takes for the pulsar to travel between two consecutive radial turning points (\textit{i.e.} two consecutive pericenter or apocenter passages). For this reason, we report our results in terms of the relativistic orbital phase
\begin{equation}
    \varphi_{\rm rel} = \frac{2\pi}{T_{\rm pre}}(t-t_p)\,,
\end{equation}
for the orbits of the eccentric toy models.

\begin{table}
\setlength{\tabcolsep}{12.5pt}
\centering
\begin{tabular}{lcccc}
\hline
Model & $a$ (AU) & $a$ (mas) & $e$   & $T$        \\   \hline
Toy 0 & 40    & 4.8      & 0   & 1.43 days  \\
Toy 1 & 175.4    & 21.1      & 0.800   & 1.1615 yr  \\
Toy 2 & 43.8     & 5.28      & 0.800  & 52.9 days  \\
Toy 3 & 5        & 0.60      & 0.786 & 2.0 days   \\ \hline
\end{tabular}
\caption{The orbital parameters (semi-major axis in both physical units and angular dimension assuming a distance of $D = 8$ kpc for the GC, eccentricity and orbital period) of the toy models used for our analysis. }
\label{tab:toy_models}
\end{table}

\begin{figure*}
    \centering
    \includegraphics[width = \textwidth]{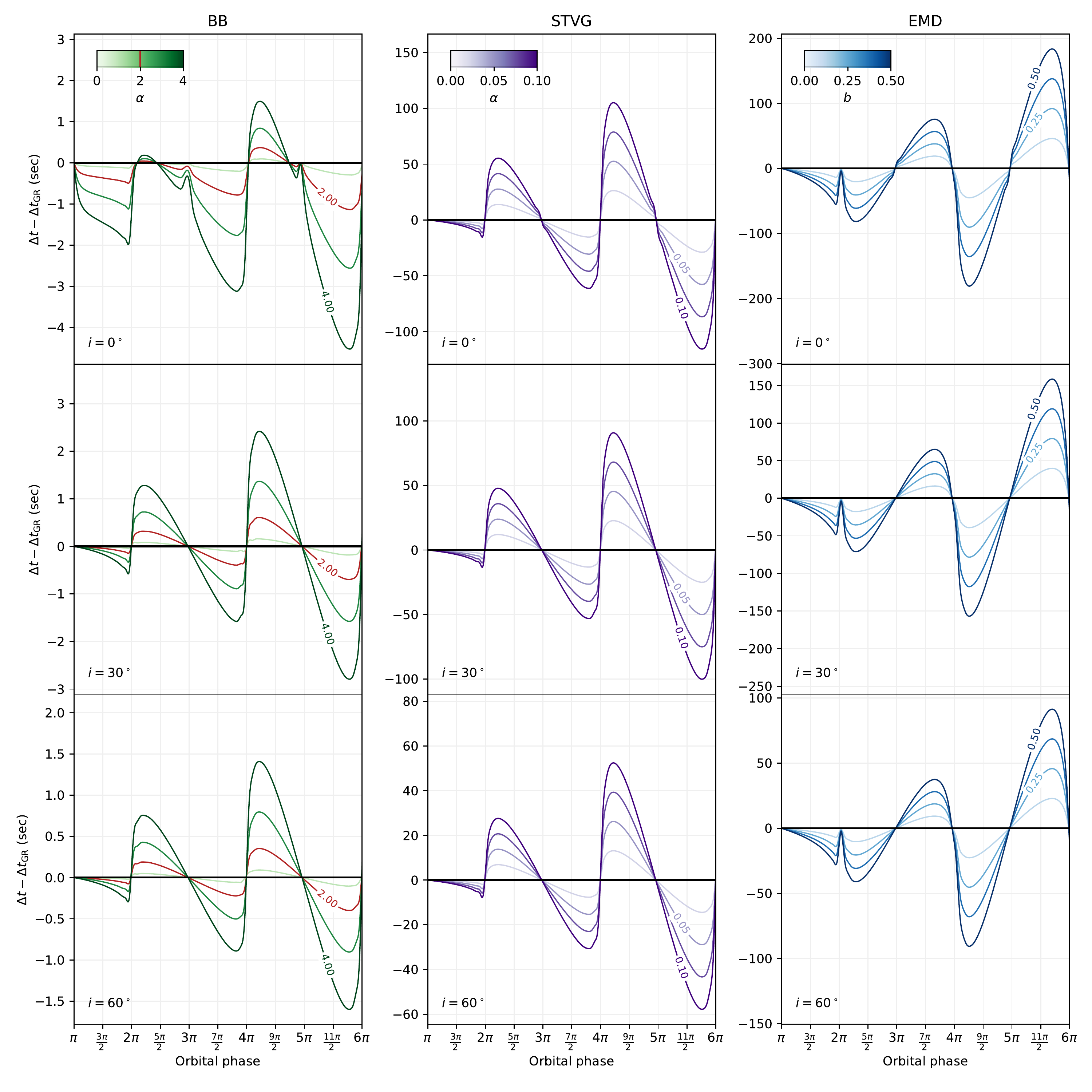}
    \caption{Differences in seconds from the Schwarzschild BH of the propagation delay in the BB (green lines, left panels), STVG (purple lines, center panels) and EMD (blue lines, right panels) models, respectively, numerically computed with our code for a central object of $M=4\times 10^6M_\odot$ and for the orbiting object Toy 1, for three inclinations of $i=0^\circ$ (top panels), $i=30^\circ$ (center panels) and $i=60^\circ$ (bottom panels) and over two and a half orbital periods (starting from apocenter $\varphi_{\rm rel} = \pi$ to $\varphi_{\rm rel} = 6\pi$). Coloured solid lines correspond to increasingly higher deviations from 
    the Schwarzschild limit obtained for different values of the theory parameter in their range of interest, as reported in the colorbars and on the plotted labels.}
    \label{fig:delays_toy1}
\end{figure*}

\begin{figure*}
    \centering
    \includegraphics[width = \textwidth]{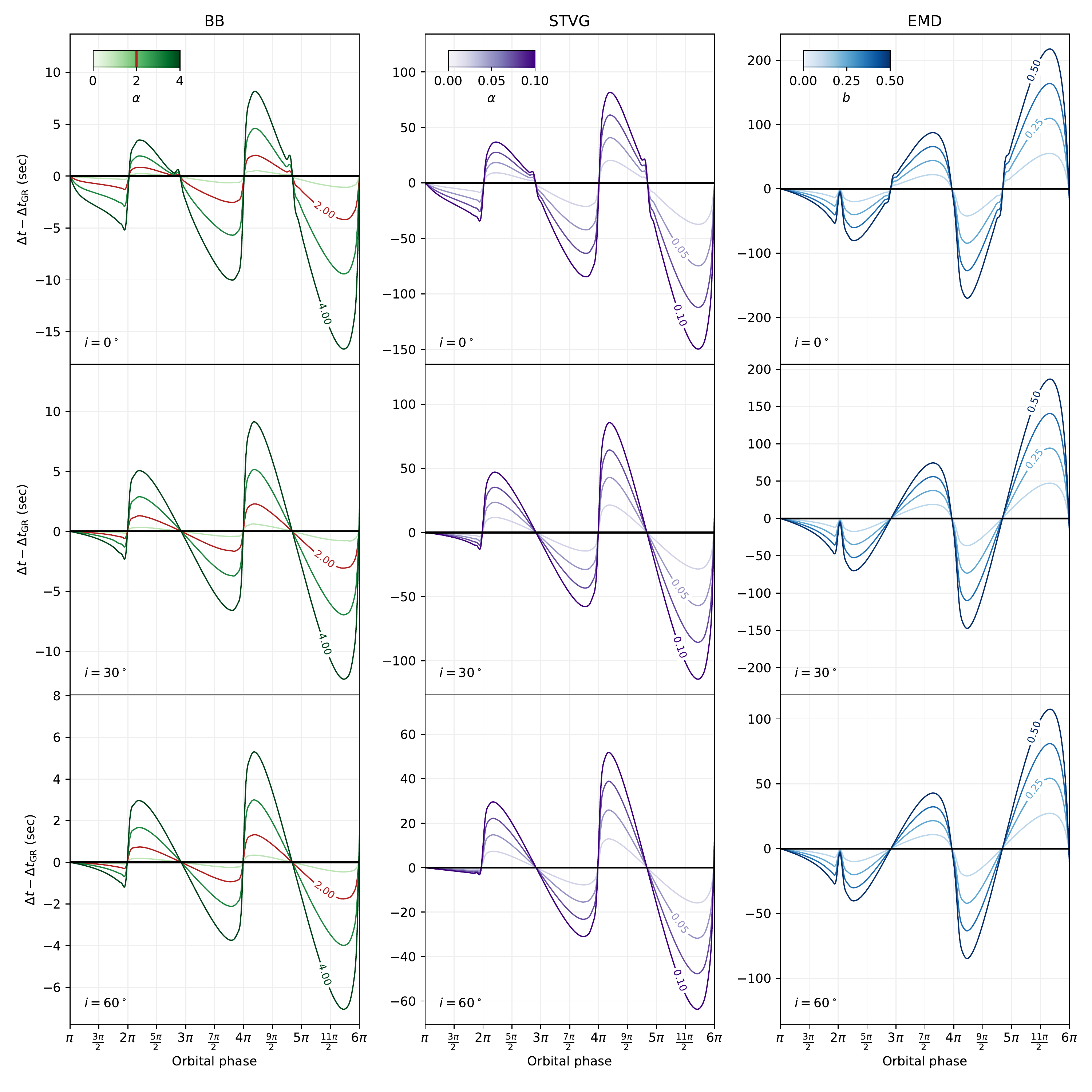}
    \caption{As in Figure \ref{fig:delays_toy1} but for the object Toy 2.}
    \label{fig:delays_toy2}
\end{figure*}

\begin{figure*}
    \centering
    \includegraphics[width = \textwidth]{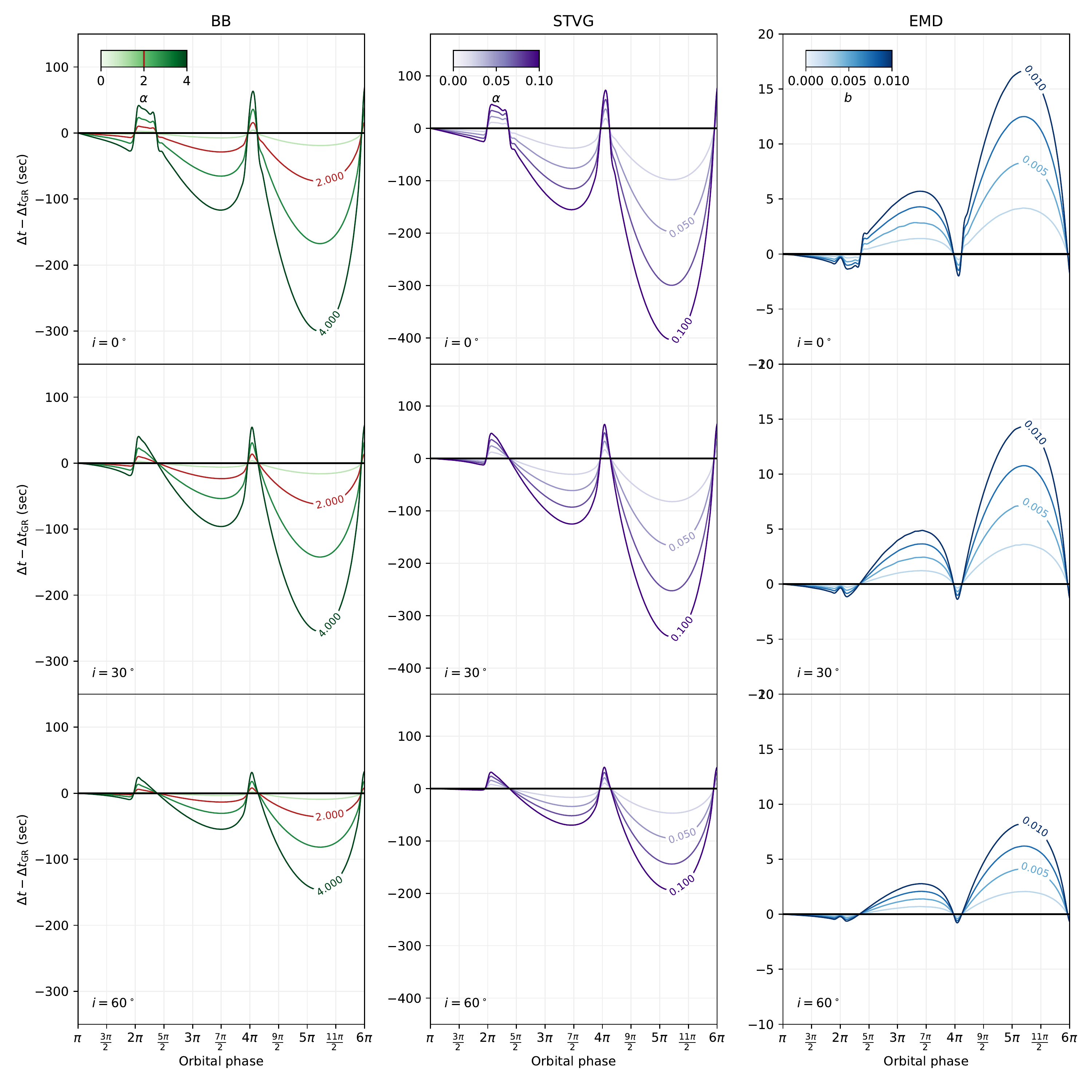}
    \caption{As in Figure \ref{fig:delays_toy1} but for the object Toy 3.}
    \label{fig:delays_toy3}
\end{figure*}

The results are reported in Figures \ref{fig:delays_toy1}, \ref{fig:delays_toy2} and \ref{fig:delays_toy3} for all the orbiting objects in Table \ref{tab:toy_models}, for the three models presented in Sec. \ref{sec:alternatives} and for the three inclinations considered. As done in the previous section, we take as a reference value the propagation time in the Schwarzschild space-time, obtained by fixing the theory parameter to zero, for all the models. The profiles shown in Figures \ref{fig:delays_toy1}, \ref{fig:delays_toy2} and \ref{fig:delays_toy3}, would thus represent the possible amplitudes of the timing residuals obtained when fitting a Schwarzschild model to the TOA data for a pulsar that orbits one of the possible alternatives to the Schwarzschild BH considered here. For each value of the theory parameter in the range of interest, we integrated the geodesic equations for the pulsar given that specific parameter value, we applied our methodology to obtain the phase-dependent photon propagation time, and then we subtracted from it the corresponding Schwarzschild propagation time. The resulting profiles, thus, encode information on both the modification to the pulsar's trajectory and to the photon paths as a function of the theory parameter. We start our orbital integration at apocenter (\textit{i.e.} at $\varphi_{\rm rel} = \pi$). Moreover, we set this as our reference point for computing the delays (\textit{i.e.} we impose that at the initial time, all the models are synchronized) so that all the profiles start at zero when $\varphi_{\rm rel}= \pi$.

A notable feature from our profiles is that for edge-on orbits ($i=0^\circ$) case, we always find an abrupt change in the propagation time when the pulsar is at superior conjunction (corresponding to values of $\varphi_{\rm rel}$ around odd multiples of $\pi$, \textit{i.e.} a configuration in which the pulsar is directly behind the central object with respect to the observer\footnote{This configuration is one example of the failure of the post-Newtonian approximation in describing the propagation delay. As the Shapiro delay calculation considers a straight-line propagation, which in this case would intersect exactly the central object, Eq. \eqref{eq:shapiro} would diverge.}). This is due to the strong curvature of photon paths to go from the emitter to the observer and to the strong impact of the Shapiro delay for such photons. Indeed, this effect is not present for inclined orbits (for which superior conjunction is never realized) and, in general, for a given orbital model and a given space-time geometry, we can always observe that for inclined orbits, the photon propagation delay always has a smaller impact, with respect to the edge-on case.

Furthermore, the departure between the different profiles from the Schwarzschild propagation time exhibits a secular increment (\textit{i.e.} the greater the number of orbital periods, the greater the departure). This is due to the increase in the orbital precession as a result of a change in the underlying space-time geometry (as in Eq. \eqref{eq:precession_STVG}, for example), which leads to an increasing departure of the spatial position along the orbit from where the photon starts.

Finally, all the mentioned effects, have an amplitude that depends on the value of the theory parameter and on the orbital properties of the object at hand. Such departures in the photon propagation time, expressed in seconds, can span several orders of magnitude (going from a few to hundreds of seconds in the most severe cases) when particularized for the SMBH at the center of the MW. Moreover, the same object orbiting different space-time geometries generates different propagation delay profiles, not only in terms of amplitude but also of functional dependence on the orbital phase, presenting peculiar features that can provide an efficient way to identify the underlying theory of gravity.

\section{Discussion and conclusions}
\label{sec:conclusions}

The possible discovery of pulsars orbiting the four-million-solar-mass SMBH in the center of the MW and the timing analysis of the radio pulses emitted by such sources would provide unique opportunities to probe the gravitational field of the GC \citep{Liu2012, Zhang2017a}. The impact of such observations on our current understanding of gravity would be unparalleled, opening up to the possibility not only to test GR in the strong-field regime at an unprecedented level of accuracy but also to falsify it against alternative theories of gravity with astounding precision \citep{Psaltis2016}.

In this work, we have presented a numerical methodology to compute the photon propagation time of light rays emitted by a pulsar undertaking a general relativistic orbit in a generic spherically symmetric space-time. We have developed a code (that we aim to publicly release near in the future) in which, in a self-consistent way, the motion of the pulsar is integrated at a geodesic level, a search algorithm for the emitter-observer problem is applied and the photon propagation time is derived directly from the geodesic equations. All GR effects are therefore automatically taken into account in our calculations (both on the orbit and on the photons themselves) in a fully relativistic fashion, \textit{i.e.} without resorting to PN approximations. Our methodology has been validated by comparing the results for a Schwarzschild space-time with the exact solution presented in \citet{Hackmann2019}. The two predictions coincide exactly within the numerical tolerance of our approach (Fig. \ref{fig:exact_vs_numerical}).

We have then turned to apply our algorithm to alternative models to the standard Schwarzschild BH paradigm, so as to highlight the generality of our approach.  In particular, we have analyzed the BB space-time, \textit{i.e.} a model in which possible alternative natures (regular BH or WH) for the central object are considered while retaining a GR framework \citep{Simpson2019}; a BH solution in STVG in which additional scalar and vectorial degrees of freedom with respect to GR are taken into account \citep{Moffat2015}; and a BH solution in EMDA, an alternative theory of gravity arising from string theory \citep{Mizuno2018}. For each of these space-time metrics, we have considered several pulsar toy models, as reported in Table \ref{tab:toy_models}, and applied our pipeline to compute the propagation delay with respect to the pure Schwarzschild case. The resulting delay residuals from GR (reported in Figs. \ref{fig:delays_toy1}, \ref{fig:delays_toy2} and \ref{fig:delays_toy3}) exhibit both single orbit features and secular deviations (\textit{i.e.} incremental over multiple orbits) that carry information on the different photon paths undertaken by light rays in the different space-times, as well as on the modification to the pulsar trajectories arising from deviations between the models at a geodesic level. Such departures span several orders of magnitude, ranging from fractions of seconds to hundreds of seconds (when particularized for Sgr A* mass), depending on the specific orbital toy model considered and on the values of the extra theory parameters in the ranges of interest. Most noticeably, the different profiles derived differ not only quantitatively, but also qualitatively, exhibiting peculiar features from space-time to space-time that would allow future pulsar timing analysis around a SMBH not only to constrain the extra theory parameters but also to completely rule out the models whose residual pattern exhibit different behaviours. Our predictions are well above the predicted timing precision of future observational facilities, like SKA \citep{Keane2015}, that aim to time GC pulsars with timing resolutions on the order of $10^{-8}\div 10^{-7}$ s \citep{Liu2011}. 

Future prospects of the present work include the extension of our approach to generic axisymmetric space-times (describing rotating gravitational sources), for whose GR counterpart, \textit{i.e.} the Kerr space-time, an exact treatment has already been formulated \citep{BenSalem2022} and has also been numerically investigated \citep{Zhang2017a, Kimpson2019}. Additionally, since long-term observations of GC pulsar might enable the detection of gravitational wave burst emission due to gravitational self-force effects \citep{Kimpson2020c}, this effect should also be taken into account. Finally, when combined with popular pulsar timing codes (\textit{e.g.} Tempo2, \citeauthor{Hobbs2006}, \citeyear{Hobbs2006}), our methodology could provide with
an extremely useful tool to estimate delays arising from the fully-relativistic treatment in the timing residual analysis for pulsars around a SMBH. This could be applied in forecasting the precision down to which extensions to GR can be constrained, when (or if) a pulsar orbiting Sgr A* will eventually be detected by future observational facilities.

\section*{Data Availability Statement}
No new data were generated or analysed in support of this research.

\section*{Acknowledgements} 
RDM acknowledges support from Consejeria de Educación de la Junta de Castilla y León. IDM acknowledges support from Grant IJCI2018-036198-I  funded by MCIN/AEI/10.13039/501100011033 and, as appropriate, by “ESF Investing in your future” or by “European Union NextGenerationEU/PRTR”. IDM and RDM also acknowledge support from the  grant PID2021-122938NB-I00 funded by MCIN/AEI/10.13039/501100011033 and by “ERDF A way of making Europe”. 


\bibliographystyle{mnras}
\bibliography{biblio} 

\bsp	
\label{lastpage}
\end{document}